\documentclass[10pt, a4paper]{article} %
\usepackage[english]{babel} %

\usepackage{microtype} %

\usepackage{amsmath,amsfonts,amsthm} %

\usepackage[svgnames]{xcolor} %

\usepackage[hang, small, labelfont=bf, up, textfont=it]{caption} %

\usepackage{booktabs} %

\usepackage{lastpage} %

\usepackage{graphicx} %

\usepackage{enumitem} %
\setlist{noitemsep} %

\usepackage{siunitx}
\usepackage{float}
\usepackage{multicol}
\usepackage{caption}
\usepackage[breaklinks]{hyperref}
\usepackage{xr}

\hypersetup{
	colorlinks,
	linkcolor={red!50!black},
	citecolor={blue!50!black},
	urlcolor={blue!80!black}
}

\usepackage{sectsty} %
\allsectionsfont{\usefont{OT1}{phv}{b}{n}} %

\usepackage{geometry} %

\usepackage[T1]{fontenc} %
\usepackage[utf8]{inputenc} %

\usepackage{XCharter} %

\usepackage{fancyhdr} %
\fancypagestyle{firstpage}{ %
	\fancyhf{}
}

\newcommand{\authorstyle}[1]{{\large\usefont{OT1}{phv}{b}{n}\color{DarkRed}#1}} %

\newcommand{\institution}[1]{{\footnotesize\usefont{OT1}{phv}{m}{sl}\color{Black}#1}} %

\usepackage{titling} %

\newcommand{\HorRule}{\noindent\color{DarkGoldenrod}\rule{\linewidth}{1pt}} %

\pretitle{
	\vspace{-50pt} %
	\HorRule\vspace{10pt} %
	\fontsize{22}{25}\usefont{OT1}{phv}{b}{n}\selectfont %
	\color{DarkRed} %
}

\posttitle{\par\vskip 15pt} %

\preauthor{\noindent} %

\postauthor{ %
	\par\HorRule %
}

\usepackage{lettrine} %
\usepackage{fix-cm}	%

\newcommand{\initial}[1]{ %
	\lettrine[lines=3,findent=4pt,nindent=0pt]{%
		\color{DarkGoldenrod}%
		{#1}%
	}{}%
}

\usepackage{xstring} %

\newcommand{\lettrineabstract}[1]{
	\StrLeft{#1}{1}[\firstletter] %
	\initial{\firstletter}\textbf{\StrGobbleLeft{#1}{1}} %
}

\usepackage[style=nature]{biblatex} %

\usepackage[autostyle=true]{csquotes} %

\DeclareSIUnit\Molar{M}

\title{Harnessing Ionic Complexity: A Modeling Approach for Hierarchical Ionic Circuit Design}

\author{
	\authorstyle{Max Tepermeister\textsuperscript{1} and Meredith N. Silberstein\textsuperscript{1,2}} %
	\newline\newline %
	\textsuperscript{1}\institution{Sibley School of Mechanical and Aerospace Engineering, Cornell University, Ithaca, NY, United States}\\ %
	\textsuperscript{2}\institution{Engineered Living Materials Institute, Cornell University, Ithaca, NY, United States}\\ %
}

\date{\vspace{-1ex}}

\begin{document}
\maketitle

\thispagestyle{firstpage} %

\begin{multicols}{2}

\lettrineabstract{Since the 1950s, soft ionic devices have evolved from individual components to an expanding library of sensors, actuators, signal transmitters, and processors. However, integrating these components into complex, multi-functional systems remains challenging due to the non-intuitive and non-linear interactions between ionic elements. In this work, we address these fundamental challenges by developing a lumped element model that enables interrogation of the physics that govern ionic circuits, as well as rapid design and optimization. Our model captures features specific to ionic charge carriers, while preserving the hierarchical design flexibility and computational efficiency of traditional circuit modeling. We demonstrate that our model can not only fit individual device behavior but also accurately predict the behavior of larger circuits formed by combining those devices. Additionally, we show how our tool utilizes the intrinsic non-linearities of ionic systems to enable novel functionality, revealing how factors such as ion enrichment, ion leakage, and polymer charge density influence performance. Finally, we present a fully ionic power supply, sensor, control system, and actuator for a soft robot that adapts its motion in response to environmental salt, illustrating the tool's potential to accelerate advancements in chemical sensing, biointerfacing, biomimetic systems, and adaptive materials.}

Soft ionic circuits---circuits that use ions as their charge carrier---promise to enable a new generation of fully soft robotics and biomedical devices that use polymers and ions instead of metals and electrons to sense stimuli, process information, and interact with their broader environment \cite{chunIontronics2015, xiongEmergingIontronicSensing2022, yangHydrogelIonotronics2018, tepermeisterSoftIonicsGoverning2022a}. Because ionic circuits are made from soft polymers they are a natural candidate to form the control systems for soft robotics. Researchers have created ionic sensors that respond to pH \cite{sheppardDesignConductimetricPH1993, richterReviewHydrogelbasedPH2008, zhaoFlexibleIonicSynaptic2018}, light \cite{mengArtificialIonChannels2014, whiteConversionVisibleLight2018, nieLightSwitchableSelfHealablePolymer2021}, and mechanical stimuli \cite{vallemSoftVariableAreaElectricalDoubleLayer2021, houFlexibleIonicDiodes2017, zhangSelfHealingAdhesiveHighly2019, jiangFlexibleTransparentAntibacterial2021, shenCutaneousIonogelMechanoreceptors2021}. Ionic signals have been processed through digital logic, in the form of \textsc{nand} \cite{hanIonicCircuitsBased2009, tybrandtLogicGatesBased2012}, \textsc{and} \cite{gabrielssonIonDiodeLogics2012, wangStretchableTransparentIonic2019, sabbaghDesigningIontronicLogic2023}, and \textsc{or} \cite{hanIonicCircuitsBased2009, wangStretchableTransparentIonic2019, sabbaghDesigningIontronicLogic2023} gates, as well as analog signal processing made from amplifiers \cite{limIontoionAmplificationOpenjunction2019, lucasIonicAmplifyingCircuits2020} and rectifiers \cite{gabrielssonFourDiodeFullWaveIonic2014, huoToughTransientIonic2023}. The low stiffness and mobile species inside ionic polymers makes them natural actuators, with longstanding work on swelling and bending based actuation \cite{kuhnReversibleDilationContraction1950, oguroBendingIonconductingPolymer1992, chengRecentProgressHydrogel2021, bhandariReviewIPMCMaterial2012}.  The bio-compatibility of many polymers and the fact that biological systems use ions and charged molecules as their signal carriers also makes ionic materials and circuits attractive for interfacing with or mimicking biological systems. Ionic circuits have delivered neurotransmitters to neurons \cite{tybrandtIonBipolarJunction2010}, monitored cells in real time \cite{yooGQuadruplexFilteredSelectiveIontoIon2023}, and delivered drugs \cite{arbringsjostromDecadeIontronicDelivery2018}. The accumulation of ions inside ionic circuits also means that iontronics are naturally suited for making devices that are able to alter their behavior in response to stimuli over time. Ionic circuits have been trained to mimic synaptic plasticity \cite{renPolyelectrolyteBilayerBasedTransparent2022} and play pong \cite{strongElectroactivePolymerHydrogels2024} using these inherent memory effects.

As ionics mature, design and simulation tools are needed that predict the response of ionic circuits and enable exploring possible circuit designs. The most common method for modeling ionic devices is discretizing a continuum theory like the Nernst-Planck-Poisson (NPP) equations using the finite element or finite volume methods \cite{doiDeformationIonicPolymer1992, wallmerspergerCoupledChemoelectromechanicalFormulation2004, bosnjakModelingCoupledElectrochemical2022a}. These continuum models have been applied to a variety of devices, including transistors \cite{volkovModelingChargeTransport2014}, diodes \cite{triandafilidiPoissonBoltzmannModeling2020, sabbaghLogicGatingLowabundance2023, tepermeisterModelingTransientBehavior2024}, and microporous ionic capacitors \cite{biesheuvelNonlinearDynamicsCapacitive2010}, and have been successful at explaining the behavior of these individual devices. However, these models are not effective as design tools when designing new or larger circuitry. First, the direct implementation of the NPP equations is extremely computationally intensive \cite{bazantDiffusechargeDynamicsElectrochemical2004, moriThreeDimensionalElectrophysiologyCable2009, henriqueNetworkModelPredict2024}, often requiring simulations that are thousands of times slower than real-time. This computational expense largely derives from the extremely fast timescales (\textless \unit{\micro\second}) and extremely small length scales (\unit{\nano\meter}) that arise in ionic solutions \cite{bazantDiffusechargeDynamicsElectrochemical2004, dreyerModelingElectrochemicalDouble2015}. Researchers have employed a number of strategies to reduce the computational complexity of these models, including only solving for the steady state solution \cite{yamamotoElectrochemicalMechanismIon2014,
	triandafilidiPoissonBoltzmannModeling2020,
	pengUnderstandingCarbonNanotubeBased2021}, using advanced time stepping strategies to deal with the timescale stiffness \cite{pughStudyNumericalStability2021, yanAdaptiveTimesteppingSchemes2021}, re-scaling Poisson's equation to enlarge the tiny lengths \cite{narayanCoupledElectrochemoelasticityApplication2021}, and assuming electroneutrality to eliminate the fast timescale altogether \cite{bazantDiffusechargeDynamicsElectrochemical2004, moriThreeDimensionalElectrophysiologyCable2009}. Each of these approaches increases the computational tractability of the simulations, but they remain slow and expensive, especially for larger ionic circuits comprising multiple ionic devices. Second, using these models requires knowledge of the geometry of the device under study. This knowledge is not a barrier to understanding the behavior of existing devices, but is a challenge when trying to design new iontronics since geometry must be drawn and meshed for every potential idea. Both the slow simulation and the requirement for geometry significantly elongate the iteration time for designing ionic circuits.

One modeling approach that resolves many of these challenges is lumped element modeling. In lumped element models, systems are viewed as a network of connected elements, with each element describing one aspect of the physics of the system and `lumping' the cumulative effect of that physics over a whole region of the device. Lumped element modeling has been applied with great success to electrical circuits \cite{macmahonCombinationsResistances1994, fosterGeometricalCircuitsElectrical1932, belevitchSummaryHistoryCircuit1962, pedersonHistoricalReviewCircuit1984,
	fairbanksReviewNonlinearTransmission2020a}, heat transfer \cite{cengelHeatTransferPractical2004a, alkhedherElectrochemicalThermalModeling2024}, multibody-mechanics \cite{ferrettiVirtualPrototypingMechatronic2004, fotsoStateArtMechatronic2012a}, and many more fields \cite{zimmerRobustModelingDirected2019a, hesterSystemsBiologyIntegrative2011}. Lumped element models resolve both of the challenges of using continuum methods for design, and also bring additional advantages. First, lumped element methods are based on graphs of connectivity between elements rather than on geometry. This means that a designer can design information and power flows between components without having to decide the physical layout of that graph. This graph based approach also enables hierarchical design: by building composite elements out of lumped elements, sub-circuits can be independently designed and then reused in larger circuits. Lumped elements also are generally much more computationally efficient than their continuum counterparts, which greatly improves the design iteration time. Because electrical lumped elements and the tools to solve circuits using them are particularly well established, there has been significant work to model ionic devices using equivalent circuits; essentially translating an ionic system to an electrical circuit with the same response \cite{finkelsteinEquivalentCircuitsRelated1963, paquetteEquivalentCircuitModel2003, nejadSystematicReviewLumpedparameter2016, 
	kimIonoelastomerJunctionsPolymer2020, haeverbekeEquivalentElectricalCircuits2022, henriqueNetworkModelPredict2024}. While this method can be effective at modeling individual devices, it fails to capture the ion identity and concentration effects fundamental to ionic conductors, and thus fails to account for many of the ways that different ionic components interact with each other. More specialized lumped element modeling has also been applied to ionic systems, with success in modeling simple electrolytes \cite{bremenDynamicModelingAqueous2022}, batteries \cite{subramaniamProperlyLumpedLithiumion2019}, and proton exchange membrane fuel cells \cite{xueSystemLevelLumpedparameter2004, FCSysModelicaLibrary2024}. These models, while useful for design of those particular devices, are specialized to a particular application, lacking generic and reusable elements. 

In this paper we present an ionic lumped element model that enables efficient, high-fidelity, design and analysis of ionic circuits operating over real-world application-relevant timescales. This model accounts for the individual ion flows and accumulations that both distinguish ionics from electronics and offer unique functional opportunities. Our approach separates the descriptions of ions, materials, elements, and devices, allowing for efficient design at a variety of abstraction levels. We create both a generic/extendable lumped element framework that determines the form and relationship between elements, and a specific implementation of many common elements. We utilize a standard lumped element modeling language, enabling easy coupling of our simulations to existing lumped element domains like traditional electrical circuit models. We show that this model captures the response of prototypical experimental ionic circuits, finding the solution much faster than real time. Finally we demonstrate design capabilities by formulating a fully ionic soft robotic circuit that senses the salinity of its environment and controls its movement accordingly. This circuit is much larger than any previously created or practical to simulate with prior techniques. Using the rapid iteration enabled by our model, we also explore some of the challenges unique to ionic circuit design and demonstrate some ways to mitigate or take advantage of these effects.

\section*{Model Formulation Enables Rapid Circuit Design via Either Material or Device Assembly}

Our ionic circuit modeling tool is enabled by ``lumped elements" that each capture one essential aspect of ionic circuit behavior in a computationally efficient manner (Figure \ref{fig:elements}). These elements can be flexibly connected together with each other and with an existing rich library of elements from other physical domains to ``build up" a functional ionic circuit. Rather than each element representing part of physical space, as is standard with continuum models like finite elements, the elements here each describe a unique part of the physics happening in ionic circuits. Elements can be combined together to create higher level models with reusable functionality; these models can then themselves be combined together into even higher level models, yielding a progressive hierarchy. This hierarchy allows circuit designers to focus on the level of abstraction most appropriate for each device. For example, devices can be constructed by explicitly specifying the layout of individual materials, or at a higher level by combining existing devices into something larger. These levels can also be mixed arbitrarily, making a device partially from low level elements and partially from higher level ones.

Ionic systems are made from a huge variety of materials, from hydrogels to ionic liquids. Many of these materials can be used with a wide range of different mobile ions. To enable hierarchically designed elements to be used with a wide variety of materials, we separate out the physics and structure represented by the elements from the material and mobile species-specific properties that compose those structures. 

\subsection*{Elements Span from Indivisible to Composite}

\begin{figure*}[t]
	\centering
	\includegraphics[width=17.8cm]{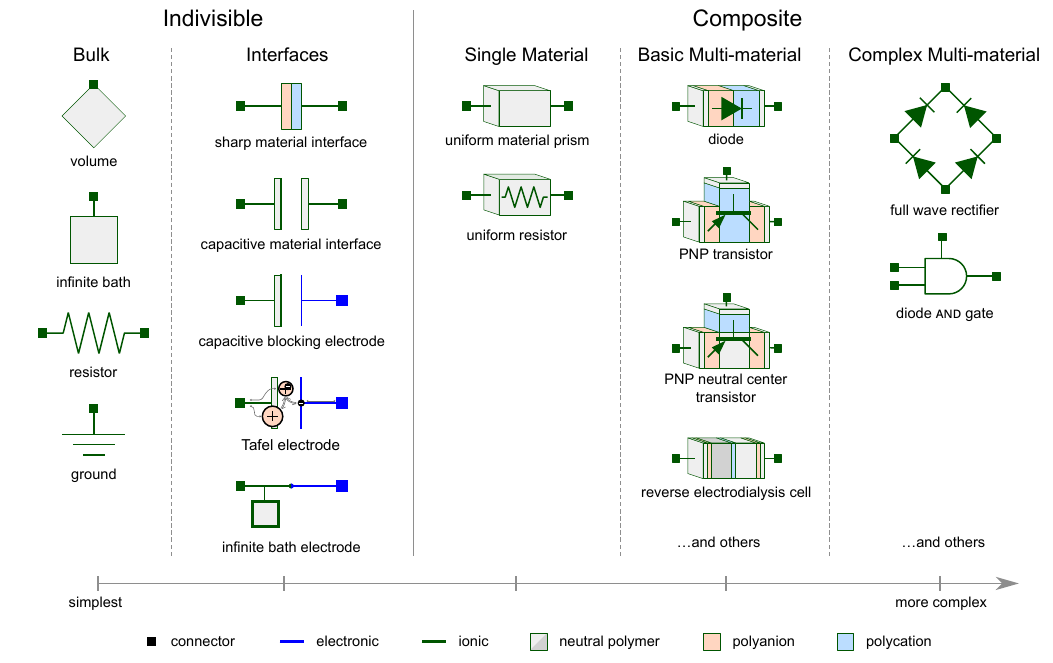}
	\caption{A categorization of the elements that we developed. These elements are separated into indivisible elements and composite elements, where the former are constructed directly from governing differential algebraic equations, and the later are constructed by combining lower complexity elements. Complexity increases from left to right. The fill colors indicate the charge of the polymer used to create it. Elements that can be made from all charges of material (for example the \textit{volume} element), are shown as neutral as a representative example. The darker grey in the electrodialysis element indicates the high salt neutral material. Green outlines indicate parts of each element that are ionic, while blue outlines indicate parts that are electrical. The solid squares on each element are connectors, and indicate where that element connects to other elements. The number of connectors varies by element.}
	\label{fig:elements}
\end{figure*}

Each indivisible element directly encodes one aspect of the physics of ionic materials (Figure \ref{fig:elements}, Indivisible). These physics are directly represented by governing differential algebraic equations. The indivisible elements either describe the bulk of materials or the interfaces between materials. Each bulk element represents a single behavior of a single piece of material. For example, the \textit{volume} element represents the concept of species accumulation and depletion. The \textit{resistor} model represents the resistance to ion motion driven by electrochemical imbalances. Interface elements are also formulated in an indivisible manner. For example, the \textit{sharp material interface} element represents an interface between two different ionic materials that are locally in chemical equilibrium, and the \textit{Tafel electrode} element represents the chemical reactions at the interface between an ionic polymer and a conductive metal electrode.

The basis for each of the indivisible elements are the NPP equations \cite{grodzinskyFieldsForcesFlows2011}. These equations describe how the electric potential and flux of mobile species, including ions and neutrally charged species, vary over space. To turn these equations into lumped elements, we must first decide what information is shared between the elements when they are connected together. This decision then constrains the formulation of the governing equations for each of the elements. Interfaces between elements in lumped element models are generally divided into `effort' variables and `flow' variables. Effort variables describe the driving forces in the system, while flow variables represent the conserved quantities that change in response to the driving forces. We choose to construct a model that uses concentrations for the state of each species and implicitly enforces electroneutrality at every connection between elements.

While any circuit can be made simply by combining indivisible elements, we develop here composite elements so that complex ionic circuits can be easily designed. The elements shown are not an exhaustive list of the possibilities of our modeling technique, but rather serve to illustrate how a wide range of devices can be created using the same building blocks and how hierarchical modeling facilitates circuit design. We categorize the composite elements based on how they are constructed (Figure \ref{fig:elements}, Composite). Each successive category is constructed exclusively from elements in the previous categories, including indivisible elements. The single material elements represent combined behaviours of single materials. For example, the \textit{uniform material prism} element combines \textit{volume} and \textit{resistor} elements together to represent a material where both ion accumulation and ion transport are at play. The basic multi-material elements take this one step further. They represent devices constructed from a few materials used in concert to achieve complex behaviors such as rectification and amplification. For example, the diode represents a combination of four material layers that allow ionic currents to easily flow in only one direction. Finally, we build complex multi-material elements that achieve functionality only possible with many individual materials and complex connectivities. For example, the \textit{diode \textsc{and} gate} uses three diodes to create a circuit that implements the logical \textsc{and} function.

\subsection*{Concentrations as the Internal States Improve Model Interpretability}

We choose to represent the state of each species using its local concentration and represent the flows between elements using molar fluxes. We choose concentrations over alternatives like molalities or mass fractions because concentrations are given in the relevant literature for ionic materials, and because using concentrations results in a direct dependence of molar flux on species mobility and electrochemical driving force, which itself depends directly on concentration (Equation \ref{eqn:elements.flux_from_force}), 

\begin{equation}\label{eqn:elements.flux_from_force}
	j_i = \frac{D_i C_i}{RT}\nabla \mu_i
\end{equation}

\noindent where $j_i$ is the flux of mobile species $i$, $\mu_i$ is the electrochemical potential of that species, $D_i$ is the diffusivity of that species, $C_i$ is the concentration of that species, $R$ is the universal gas constant, and $T$ is the temperature.

An alternative natural state variable would be the electrochemical potential since it is the power conjugate of the species flux. We choose to use concentrations over electrochemical potentials because electrochemical potentials are hard to measure experimentally. Using concentrations also eases the direct implementation of chemical potential models that are written in terms of the concentrations (\textit{e.g.} the Pitzer or Debye-H{\"u}ckel models \cite{debyeTheoryElectrolytesFreezing2019, pessoafilhoExtensionPitzerEquation2008}) without having to solve the inverse problem from electrochemical potential to concentration.

\subsection*{Implicit Electroneutrality Creates Local Balance}

The second important modeling choice is to implicitly apply an electroneutrality constraint at every connection between elements. This choice greatly reduces the computational cost of solving large models, improves the flexibility with which elements can be connected to each other, and removes pitfalls with circuits that contain loops.

The NPP equations, upon which our model is based, allow for regions of space to have a net charge. However, this so called `space charge' is generally a very high energy state because electrostatic interactions between charged regions are very strong compared to diffusive forces. This energy cost thus constrains the time and length scales over which a region can be charged \cite[p.53,73]{grodzinskyFieldsForcesFlows2011}. The length scale for this neutralization is known as the Debye length, and for many ionic systems is on the order of nanometers. This small length scale implies a similarly small timescale to reach an electrically neutral state at distances larger than the Debye length from an interface, typically on the order of nanoseconds to microseconds \cite{ grodzinskyFieldsForcesFlows2011, bazantDiffusechargeDynamicsElectrochemical2004}. The systems that we are interested in modeling are both orders of magnitude larger (microns to millimeters) and have transient responses that are orders of magnitude slower (milliseconds to hours) than the electroneutrality length and timescales. Much of the significant computational cost in directly implementing the NPP equations comes from resolving the fine spatial and time resolutions created by electro-non-neutrality \cite{bazantDiffusechargeDynamicsElectrochemical2004, moriThreeDimensionalElectrophysiologyCable2009}. To significantly simplify the implementation of our model, we replace this extremely fast timescale with a constraint that our elements are always electrically neutral at their connections with other elements. The bulk elements are also electro-neutral internally, while some of the interface elements have internal charge in order to model effects like double layer capacitance. For these internal charges we also formulate this charge in terms of equilibria instead of dynamics in order to remove the equilibration timescale (Equation \ref{eqn:elements.electrode.blocking-1}). Equation \ref{eqn:elements.electroneutrality_condition} expresses electroneutrality, where $z_i$ is the elementary charge of species $i$, and $\rho_q$ is the non-mobile net fixed charge density in the element's material.

\begin{equation}\label{eqn:elements.electroneutrality_condition}
	\sum_i F C_i z_i + \rho_q = 0
\end{equation}

We implement this electroneutrality condition implicitly at the connections between elements by removing one charged species from the information shared between elements. We then calculate this concentration based on the electroneutrality condition (Equation \ref{eqn:elements.electroneutrality_condition}) whenever an element needs it. The final information that is shared between elements is thus $n-1$ mobile species concentrations and the electrical potential as effort variables and $n$ mobile species fluxes as flow variables. This implicit construction ensures that each element is `locally balanced', \textit{i.e.,} when given information about all of the flow variables at the connections to other elements, all the effort variables can be calculated, and vice versa. Local balance enables any elements to be connected with any other and be globally balanced \cite{modelicaassociationModelicaReferenceBalancedModel2024}. 

Our use of implicit electroneutrality does require that the description of an interface between two materials cannot be split into two elements, since interfaces are often charged. For example, an interface with coupled chemical reactions and capacitive charging effects must be described as a single element since information about the non-electroneutral coupling between the different physics cannot be shared over our defined interface.

\subsection*{Separate Material and Element Models Enable Flexible Design}

Each indivisible element represents physical laws separate from the choice of material or the choice of chemical species that flow through that material. This separation allows the same element to be used in many devices without having to create material or species specific elements. For example, the \textit{sharp material interface} represents a local electrochemical equilibrium at the interface between any two materials with any number of mobile species. 

To enable this separation, we define a hierarchy of material and species properties: each element is made of one or more materials and each material has an associated set of mobile species. The states of those materials are always defined at each of the element's connectors and sometimes also internally. The properties of each material's mobile species and the thermodynamic state of that material are used to calculate the properties of each material, which in turn are used by elements to implement their governing laws. To connect the connectors of two different elements together, those connectors must share a material model and set of mobile species.

The most basic definition of a mobile species requires only the constant $z_i$, the charge of a single molecule of that mobile species. Species can define more constants, such as the radius of their hydration shell in water or their polarizability, which can then be used by materials models that depend on those values. The set of parameters for species $i$ is denoted $Y_i$. 

Each material model defines a thermodynamic state that fully describes the state of the material. For all of the materials used in this paper, the thermodynamic state for each material is $X = \begin{bmatrix} T & C_1 & \hdots & C_n \end{bmatrix}$, where $T$ is the material temperature and $C_i$ is the concentration of the $i$th mobile species out of $n$ total species. If material properties depended on other thermodynamic variables like the pressure, these would be included in this thermodynamic state. The values of this thermodynamic state must either be a specified constant or calculated from information shared between elements. Each material model then defines fixed charge density ($\rho_q(X, Y_i, \dots Y_n)$), individual species diffusivity ($D_i(X, Y_i, \dots Y_n)$) and activity coefficients ($\gamma_i(X, Y_i, \dots Y_n)$) as constants or functions of its thermodynamic state and species parameters.

Elements use these material properties in their governing equations. For example, the \textit{sharp material interface} is made from two different materials that share a set of mobile species, with one material chosen for each connector. The activity coefficients of each of the mobile species in each material are calculated based on the species properties and concentrations and are then used in enforcing chemical equilibrium between the two sides (Equation \ref{eqn:elements.sharp_interface.equilibrium}). The \textit{capacitive material interface} is similar, with two different materials representing each side of the interface. However, unlike the \textit{sharp material interface}, each material can have its own set of mobile species because they are unable to mix; passing current only through an internal electric field. This element also has internal states representing species accumulation on each capacitor plate, which is not electrically neutral (Equation \ref{eqn:elements.electrode.non-blocking-1}).

\section*{Lumped Models Match Experimental Circuit Performance and Predict Combined Behavior} 

\begin{figure*}[t]
\centering
\includegraphics[width = 17.8cm]{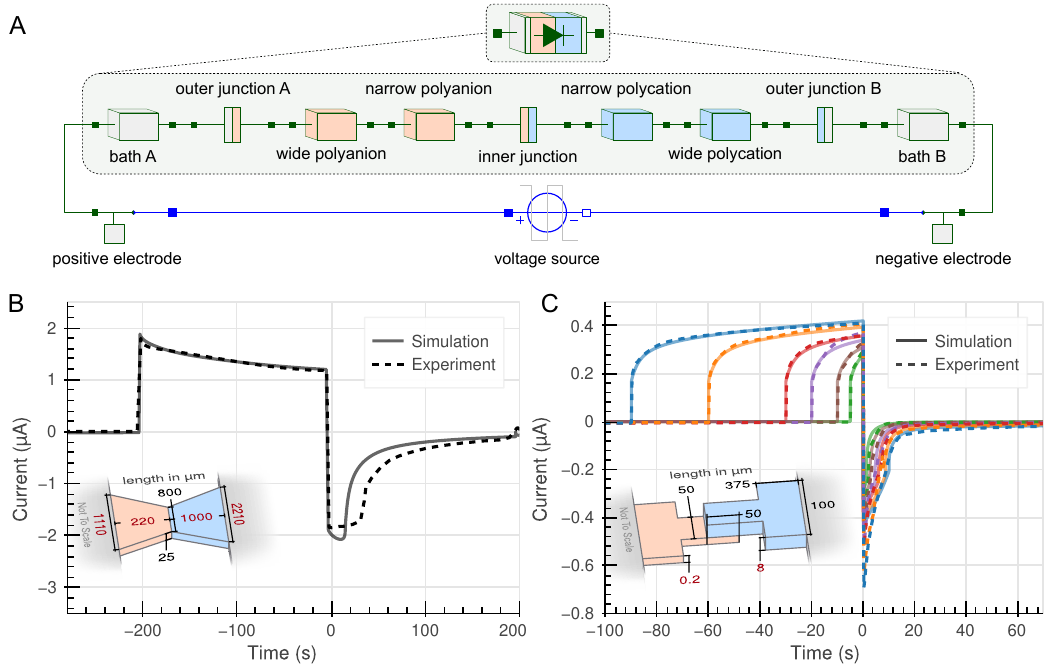}
\caption{Structure and behavior of single polyelectrolyte diodes. A) The ionic circuit. The small green box shows the symbol used for ionic diodes, and the larger green box shows the elements that compose that diode. Elements outside of the green box represent the boundary conditions and electrical voltage source used to apply voltages and measure currents of the diode. B) Simulated (solid) and experimental (dashed) current vs time responses of the diode created by Sabbagh et al. to an applied square wave voltage \cite{sabbaghDesigningIontronicLogic2023}. The voltage changes from zero to positive (forward bias) at \qty{-200}{\second} and from positive to negative (reverse bias) at \qty{0}{\second}. The inlay in the lower left shows the original diode geometry with labeled lengths in \unit{\micro\meter}. Note that this diagram is not drawn to scale. The black lengths on the inlay are estimated from the Sabbagh et al \cite{sabbaghDesigningIontronicLogic2023}, while the lengths in red are fit to match the experimental data. C) Simulated (solid) and experimental (dashed) current vs time responses of the diode by Gabrielsson et al. to an applied square wave voltage \cite{gabrielssonPolyphosphoniumBasedBipolar2013}. The colors correspond to the time spent in forward bias (\qty{5}{\second}, \qty{10}{\second}, \qty{20}{\second}, \qty{30}{\second}, \qty{60}{\second}, \& \qty{90}{\second}). All experiments transition from forward to reverse bias at \qty{0}{\second}. The inlay in the lower left shows the original diode geometry, not to scale, with annotated length colors as in B.}
\label{fig:literature.single_diode_comparison}
\end{figure*}

At the core of all ionic circuits are interfaces between different materials. These interfaces---when formed between materials with opposite fixed charges---act like one way valves, only allowing ions to flow easily in one direction. This structure is known as an ionic diode \cite{kooIonicCurrentDevices2013}. The simplicity of constructing an ionic diode, and the non-linearity that they provide, means that these diodes are at the core of most existing ionic circuitry, enabling everything from programmed neurotransmitter delivery to temperature sensing to digital logic \cite{gabrielssonIonDiodeLogics2012, duIonicDiodebasedSelfpowered2022, hanIonicCircuitsBased2009}. This makes them a natural device to demonstrate the success of our hierarchical method in breaking up ionic circuits into efficient, composable pieces. We fit a simple model to experimental results for two different single ionic diodes and then use this model to accurately predict the performance of larger circuits built by assembling multiple diodes together. 

\subsection*{Matching Single Diodes}

We build a generic diode model (Figure \ref{fig:literature.single_diode_comparison}A) that matches the transient current responses of two experimental diodes from the literature and populate this model with known experimental parameters and unknown parameters fit from the experimental results. In the process we gain insight into what creates their very different response shapes (Figure \ref{fig:literature.single_diode_comparison}B,C).

The polyelectrolyte ionic diodes we model are formed from four layers of stacked materials. These layers are a neutral layer, a polyanion (negatively charged), a polycation (positively charged), and a second neutral layer (Figure \ref{fig:literature.single_diode_comparison}A). Each layer of material in the diode is represented by one or two \textit{uniform material prisms} (SI Appendix Section \ref{sec:si.uniform_material_prism}). Distinct materials are separated by \textit{sharp material interfaces}. The \textit{uniform material prism} models both the species accumulation and the resistance to species motion of one material. The \textit{sharp material interface} captures the electrochemical equilibrium between two materials in contact with each other. 

The fundamental forward bias response curve of this type of ionic diode is broken into two stages \cite{tepermeisterModelingTransientBehavior2024}. 
In the first stage the current falls as the outer bath--polyelectrolyte junctions are depleted of ions. In the second stage the current rises as ions from the inner junction cross over the polyelectrolytes and reach the outer junctions. The reverse bias response is also broken into two stages, with a large current as the inner polyanion--polycation junction is depleted, and then a slowly decaying current as the rest of the polyelectrolyte volume is depleted.

The first experimental diode that we model (Figure \ref{fig:literature.single_diode_comparison}B) is a polymer gel diode constructed on a microfluidic chip by Sabbagh et al. (henceforth referred to as the trapezoidal diode) \cite{sabbaghDesigningIontronicLogic2023}. The response of this trapezoidal diode is dominated by stage one depletion: the ions which cross the inner junction do not have time to reach the outer junctions because the width of the polyelectrolyte regions ($\approx$\qty{500}{\micro\meter}) is large compared to the polyelectrolyte diffusivities ($\approx$ \qty{3e-10}{\meter\squared\per\second}) and fast cycle time (\qty{200}{\second}).

The second diode that we model (Figure \ref{fig:literature.single_diode_comparison}C) is a thin film diode constructed by Gabrielsson et al. (hereafter referred to as the thin-film diode) \cite{gabrielssonPolyphosphoniumBasedBipolar2013}. By fitting the model to this second diode, we learn why it behaves so differently from the trapezoidal diode in forward bias, even though its geometry and construction is similar. The thin film diode is dominated by stage two enrichment because its material diffusivities are large compared to the polyelectrolyte width. This behavior is enhanced by the large driving force (\qty{4}{\volt}), which leads to a large degree of ion enrichment in the polyelectrolytes \cite{tepermeisterModelingTransientBehavior2024}. This enrichment is additionally separated into two distinct timescales: the current initially increases rapidly and then instead of leveling off continues increasing more slowly. These two timescales are created by the different responses of the polyanion and polycation side of the diode. The polyanion (pEDOT:PSS) is both quite conductive and very thin (fit as $\approx$ \qty{200}{\nano\meter}), whereas the polyanion (pVBPPh3) is both not very conductive and thicker (fit as $\approx$ \qty{8}{\micro\meter}). The conductances of the two polyelectrolytes are relatively similar so neither dominates the current response, but the polyanion has a stage two timescale about 2 orders of magnitude faster than the polycation.

The model fit with realistic parameters captures the response of both diodes to a square wave voltage (Figure \ref{fig:literature.single_diode_comparison}B,C). We match the forward bias timescales and magnitudes well, and miss the magnitude of the initial reverse bias current by approximately 50\% for both diodes. For the trapezoidal diode, we overestimate this current, which is partially explained by our assumption that the central interface is sharp, whereas in the experimental system, its likely diffuse. For the thin film diode we underestimate the current, which is partially caused by approximating the diode as one dimensional. This approximation eliminates the \qty{50}{\micro\meter} material overlap at the inner junction (Figure \ref{fig:literature.single_diode_comparison}C inlay) and reduces how fast ions can re-cross the interface when the voltage flips. Additionally, by choosing to construct our model with only a few elements, we smooth away information about the fastest timescales for both diodes. A full description of how we customize the circuit model to each diode is provided in SI Appendix section \ref{sec:si.single_diodes}; a full table of model parameters is given in SI Appendix Table \ref{tbl:si.diode_parameters}. For both of these diodes, and all the simulations we present in this work, our model is solved much faster than real time (SI Figure \ref{tbl:si.computational_cost})

\subsection*{Predicting Multi Diode Transients}

\begin{figure*}[!t]
\centering
\includegraphics[width = 17.8cm]{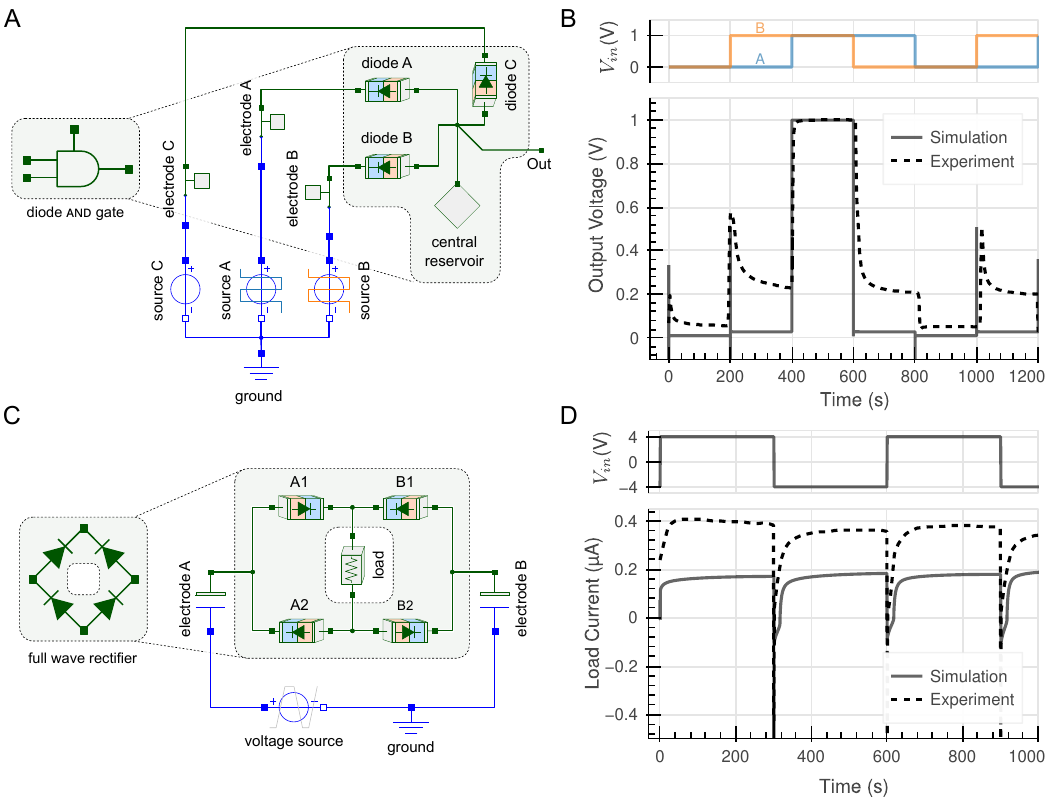}
\caption{Structure and transient behavior of multi-diode circuits. A) Circuit diagram for an \textsc{and} gate constructed from the trapezoid ionic diodes of Sabbagh et al. \cite{sabbaghDesigningIontronicLogic2023} The element is shown on the left and the experimental circuit is shown on the right with the elements that comprise the gate highlighted in the green region. The other elements model the applied boundary conditions and electrical voltage sources B) Simulated (solid) and experimental (dashed) voltage at the `out' electrode vs time for the \textsc{and} gate (bottom) subject to changing input voltages (top). C) Circuit diagram for a full wave rectifier constructed from the thin-film ionic diodes of Gabrielsson et al. \cite{gabrielssonFourDiodeFullWaveIonic2014} The elements that comprise just the rectifier are similarly highlighted in green, with the rest of the elements describing the boundary conditions and voltage source. The electrode capacitance and area is taken from the Gabrielsson et al. \cite{gabrielssonFourDiodeFullWaveIonic2014}, and the diodes are the same as fit in Figure \ref{fig:literature.single_diode_comparison}C. D) Simulated (solid) and experimental (dashed) currents vs time for the full wave rectifier (bottom) subject to an applied square wave voltage (top).}
\label{fig:literature.multi_diode_circuits}
\end{figure*}

One of the strengths of the circuit modeling technique is in reusing models when creating higher level circuits. We can therefore design circuits with more advanced functionality by combining the functionalities of simpler circuits. We demonstrate this strength by combining multiple diode elements together to form a \textit{diode \textsc{and} gate} and a \textit{full wave rectifier} that match the behavior of experimentally combined diodes without additional fitting. 

One of the most developed applications of ionic diodes is as the basis for digital logic \cite{hanIonicCircuitsBased2009,
tybrandtLogicGatesBased2012,
gabrielssonIonDiodeLogics2012,
wangStretchableTransparentIonic2019,
sabbaghDesigningIontronicLogic2023,
hanIonicCircuitsBased2009,
wangStretchableTransparentIonic2019,
sabbaghDesigningIontronicLogic2023}. Digital logic is formed from logic gates that implement basic operations such as \textsc{and} and \textsc{or}. We combine three of the modeled thin-film diodes into a logical \textsc{and} gate (Figure \ref{fig:literature.multi_diode_circuits}A) that matches the structure of Sabbagh et al.'s implementation of the same circuit \cite{sabbaghDesigningIontronicLogic2023}. In this circuit, two diodes (A and B) pull the central output to a low voltage when either of the input voltages is low. Diode C is always reverse biased and is used as a pull-up resistor to pull the output voltage high when neither input is pulling it low. We predict the response of this circuit to input voltages that step through each possible input combination in sequence (Figure \ref{fig:literature.multi_diode_circuits}B). We successfully predict the logical operation of the circuit---the output is at a high voltage only when both inputs are---and the small increase in output voltage when only one of the inputs is on. We also predict the voltage output value when both inputs are high, but predict a lower output voltage than the experiments when either or both inputs are low. This an be largely explained by leakages in the circuit that bypass the junctions. When we incorporate those leakages into the circuit representation, we recover most of the difference between experiment and simulation results (SI Appendix Figure \ref{fig:si.logic_gate_leakage}).

In addition to digital logic, diodes play a key role in some ionic power supplies. Supplying continuous current to ionic circuits using electrical power supplies is challenging because you run out of ions at the interface between the electrical supply and the ionic circuit. This issue can be resolved by instead supplying alternating electrical current and then converting it into constant current within the ionic circuit. This conversion can be done with an arrangement of four diodes in a \textit{full wave rectifier} (Figure \ref{fig:literature.multi_diode_circuits}C). Our model of this rectifier circuit matches Gabrielsson et al.'s experimental implementation, which combines 4 thin film diodes together with two capacitive electrodes and a resistive load  \cite{gabrielssonFourDiodeFullWaveIonic2014}. A square wave voltage is applied to the two electrodes, and a relatively constant current flows through the resistive load attached to the output of the rectifier. With no modification of the thin-film diode parameters used in the single diode comparison, the simulation result matches the timescales and the shape of the load current (Figure \ref{fig:literature.multi_diode_circuits}D). The magnitude of predicted currents is roughly half that of the experiments. This difference is likely due to batch-to-batch variation of the thin-film diodes as evident from the difference in single diode responses between the single-diode paper \cite{gabrielssonPolyphosphoniumBasedBipolar2013} and the multi-diode-rectifier paper \cite{gabrielssonFourDiodeFullWaveIonic2014}.

\section*{Applying the Ionic Circuit Modeling Framework to Design a Soft Robot}

\begin{figure*}[!t]
\centering
\includegraphics[width = 17.8cm]{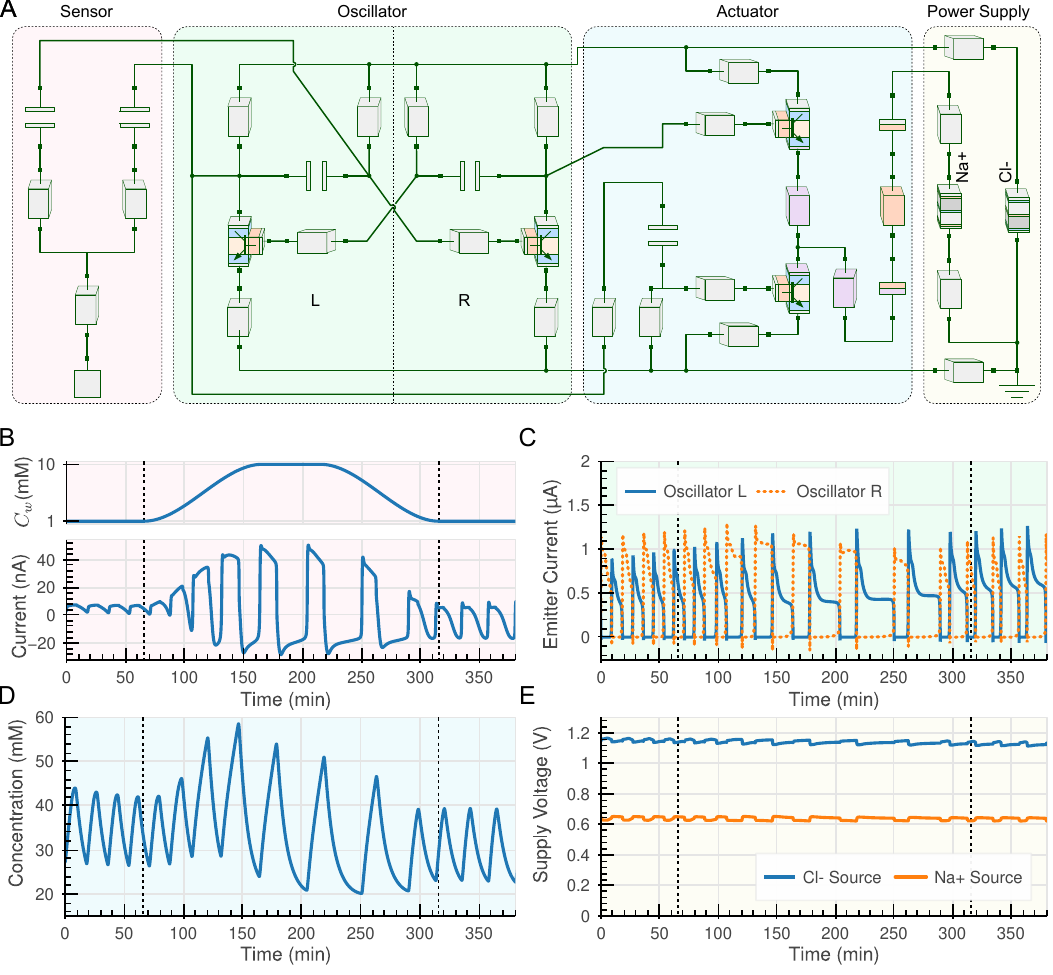}
\caption{Structure and transient behavior of powered soft robot controller circuit. A) The elements that form the circuit, broken up into the four modules by color. From left to right these are the sensor (pink), oscillator (green), actuator (blue), and power supply (yellow). The purple elements in the actuator module represent the differentially swelling material bi-layer. The other colors match those in Figure \ref{fig:elements}. B) The environmental concentration of salt (upper) and the sensor current (lower) vs time. The environmental concentration profile is chosen as an input to the system. The vertical bars are shared by all the plots, and correspond to the beginning and end of the elevated environmental salt concentration C) The current flowing through the emitters of the two oscillator transistors over time. The frequency of oscillation is controlled by the connected sensor. D) The concentration of salt in the middle of the actuator material over time, which will drive the degree of actuation. E) The voltage across the sodium and chlorine voltage sources over time. The voltage remains relatively stable even as the rest of the circuit draws variable current.
}
\label{fig:design.oscillator_circuit}
\end{figure*}

Next, we design the complete control and power system for a fully ionic soft robot that is powered by the energy of mixing salty and fresh water, senses the pH or salinity of its environment, and spends more time in high salinity environments by controlling the oscillation frequency of a bending actuator. The soft robot is composed of four separate sub-circuits -- power supply, actuator, oscillator, sensor -- each responsible for a particular aspect of its behavior (Figure \ref{fig:design.oscillator_circuit}A). The sensor detects salinity, shifting the frequency of the oscillator, which then shifts the frequency and amplitude of a bending actuator. This circuit demonstrates the power of our design method for simulating ionic devices larger than any previously designed, as well as reveals design principles that differ from traditional electronics and are necessary for designing functional ionic circuits.

To spend more time in environments that are higher salinity, the robot must sense changes in the salinity around it. To do so, it uses a sensor module that detects the conductivity of the external world using a small piece of neutral hydrogel exposed to the external environment (Figure \ref{fig:design.oscillator_circuit}B). This sensor is isolated from the rest of the control circuitry using ionic capacitors so that external salinity changes do not change the carefully tuned resistances of the internal wiring.

The oscillator module processes the sensor measurements to control the actuator cycle with an alternating signal. The oscillator is designed after a classical electronic astable multivibrator. It consists of two NPN transistors, adapted from existing ionic transistors \cite{tybrandtIonBipolarJunction2010} (SI Appendix Figure \ref{fig:si.transistor}A,B), that are connected with negative feedback to each other. When one transistor begins to turn on, it turns the other transistor off, which further enforces the opening of the first, resulting in a net positive feedback loop. A capacitor combined with a leakage resistor forces the off state of each transistor to be unstable (if left alone it will gradually turn on), which forces the entire circuit to oscillate (Figure \ref{fig:design.oscillator_circuit}C). The higher the conductivity of the sensor, the larger the capacitance in parallel with the leakage resistor, and the longer the oscillator stays in one of its two states, which lengthens the oscillation period of the oscillator. 

In order to move around its environment, the robot uses an ionic bending actuator. This actuator is composed of a bi-layer of neutral polymers that differentially swell in response to changing internal salt concentrations, causing it to bend \cite{chengRecentProgressHydrogel2021}(purple material in Figure \ref{fig:design.oscillator_circuit}A). The salt concentration is controlled by two more ionic transistors (SI Appendix Figure \ref{fig:si.transistor}A,C). When the lower transistor is on, the salt concentration inside the actuator increases: chlorine ions are forced into the actuator and are met by sodium ions from the sodium power supply. When the upper transistor is on, the salt concentration in the actuator decreases: chlorine ions are removed from the polymer and the sodium ions are forced back into the sodium power supply. The oscillator alternates between powering these two transistors, which leads to a cyclical bending of the actuator via cyclical enrichment and depletion of salt in the actuator (Figure \ref{fig:design.oscillator_circuit}D and SI Appendix Figure \ref{fig:si.oscillator_circuit.currents}). The cycle period control enables the environmental responsiveness of the robot. In regions of high salinity, the robot will move slowly because a long period of time passes between each bending of the actuator (SI Appendix Figure \ref{fig:si.oscillator_circuit.sensor_sensitivity}). In low salinity environments, the robot moves much faster, moving quickly around its environment until it once again is in a high salinity region.

The entire robot is powered by a fully ionic power supply module. This supply combines salty and fresh water in a reverse electrodialysis cell and uses the energy of mixing to drive chlorine ions through the rest of the circuit. The cell is formed from thin layers of Nafion and FAS membranes that separate 14 alternating chambers of high-salt and low-salt water (SI Appendix Figure \ref{fig:si.electrodialysis}), generating about \qty{1.2}{\volt} of potential. The majority of the circuit is powered by the flow of these chlorine ions. The power supply also has a second source (with 7 chambers and about \qty{0.6}{\volt}) that supplies sodium ions to the actuator, since the actuator requires a source of both positive and negative ions. The voltage produced by these two power supplies is shown over time in Figure \ref{fig:design.oscillator_circuit}E. These power supplies are able to maintain relatively steady voltage throughout the actuator cycle and independently of actuation frequency.

\section*{Discussion}

Ionic circuits must be constructed differently from electronic circuits. These differences present challenges, but also opportunities for additional functions and device performance improvement versus the current state-of-the-art. While some of these individual effects have been previously reported in the literature (\textit{e.g.} the enrichment of neutral wires in Doi et al. \cite{doiDeformationIonicPolymer1992} and the role of transistor base polymer charge in Volkov et al. \cite{volkovModelingChargeTransport2014}), their effect on circuit design has not been discussed as far as we are aware.

One of the organizing principles of designing ionic circuits is that each type of charge carrier is not interchangeable with another, \textit{i.e.} current carried by sodium ions in one part of circuit cannot easily be turned into current carried by chlorine ions in a different part of the circuit. The consequences of this principle lead directly to two rules for circuit design. First, in order for a part of a circuit to operate in steady state (DC), there must be a continuous path for each charge carrier to make a full loop. This is stronger than the constraint on electronic circuits, where only the current must make a full loop, and can be carried by different charge carriers in different regions. If this principle is not obeyed, then ions will either accumulate or be depleted in some part of the circuit (by the divergence theorem and species conservation). This accumulation/depletion will change the local conductivity of that region of the circuit, and if unintended may eventually stop the circuit from working.

Second, the conductivity of resistors/wires will change with current flow where those wires connect to polyelectrolytes, even when the overall ion flow is balanced. The direction of this change can be predicted by examining the materials connected to the resistor. When a neutral material is touching a polyelectrolyte, the neutral material will increase in conductivity when the current flowing into the polyelectrolyte material has the same sign as the fixed charge in the polyelectrolyte (SI Appendix Figure \ref{fig:si.wire_concentration_changes}). For example, a positive current flowing into a polycation will increase the conductivity of the attached neutral polymer. This occurs because in steady state the same anionic current must flow in the neutral polymer and the polycation. In the neutral polymer, whatever voltage drives this flow of anions must push equally on the cations, but these cations are not able to easily flow in the polycation. Thus, these cations build up near the neutral--polycation interface until a concentration gradient of cations is created that cancels most of the electrical driving force. This effect is most prominent when a non-negligible voltage drop occurs over the neutral wire and must be accounted for when designing the geometry of an ionic circuit. 

When we designed the soft robot power and control modules, we both compensated for and used the effects of both of these principles. All of the main currents are carried by chlorine ions. The reverse electrodialysis cell was arranged with polycationic membranes facing its outside such that it pushes mainly chlorine ions around the external circuit, while sodium ions mainly flow inside the cell itself. To match this flow of chlorine ions from the power supply in the rest of the circuit, we are restricted to using NPN transistors in the oscillator and actuator. A full visualization of the flow of sodium and chlorine in both oscillator states is given in SI Appendix Figure \ref{fig:si.oscillator_circuit.currents}.

However, we deliberately violate this carrier loop rule when constructing the actuator. The actuator works by enriching and depleting a gel of ions. To achieve this effect, we connect one side of the actuator to a power source that mainly supplies sodium ions, while the other side is connected to the source that mainly supplies chlorine ions. When these currents meet in the actuator, the positive ions and negative ions cannot easily pass through, and so they instead enrich the actuator gel. When the current is reversed, the opposite happens and ions are depleted from the region. As an additional consequence, current only flows until all of the ions in the actuator are depleted. This enrichment and depletion behavior is also used in the center of each transistor; only negative ions flow easily through the collector and emitter, while positive ions flow through the base. When the flow of both of these carriers points toward the center, it increases the transistor conductivity, while when they flow away from the center it decreases the conductivity, allowing the transistor to control currents. 

We also compensate for the second effect in designing the robotic circuit. Thin neutral polymer wires connect to polycations at the top of each of the transistors. We make those wires smaller than the size required by equilibrium resistance of those wires would indicate. Then, when current flows in the circuit, this decreased size compensates for the increase in conductivity as the concentration inside increases. The opposite case occurs at the bottom of each of the transistors, and so those wires are increased in size to compensate.

While these ion loop constraints and changes in conductivity pose a challenge to directly copying existing electronic circuit designs, it opens the door to novel designs and devices not easily possible with electronic circuits. For example, this behavior naturally creates a circuit memory effect where the conductivity of a material can depend on the history of current that has flowed through it. This memory can be leveraged for intrinsic neuromorphic computation or circuit adaption \cite{strongElectroactivePolymerHydrogels2024,
renPolyelectrolyteBilayerBasedTransparent2022}. Our tool provides a simple way to design with these effects.

Its also important to note that control and variation of the fixed charge density of the polyelectrolytes has the potential to improve the performance of ionic diodes and transistors, and therefore circuits. In order for large currents to flow through ionic devices, the wires that connect those devices together cannot be too resistive. If the fixed charge density of the polyelectrolytes can be increased, the salt content in the neutral connecting wires can also be increased without increasing the leakage or other unwanted behaviors. For transistors though, larger fixed charge densities are not always better, and there is a potential for large device improvements through better control of the charge density in different regions. When adapting the experimental transistors to work for the proposed soft robotic circuit, we added a slightly charged polyelectrolyte in the center region, which significantly reduced the transistor leakage. In silicon transistors, the fixed charge density (doping) of each region is very tightly controlled \cite[p. 320]{szePhysicsSemiconductorDevices2007}, and we expect that similar controls over each region in ionic transistors would lead to improvements in both leakage and gain.

\section*{Conclusions}

In this work, we conceptualized an open-source modeling tool designed to advance the development of a wide range of soft ionics-based technologies through its versatile capabilities. By seamlessly interfacing with mature electrical circuit modeling tools, our model leverages the strengths of existing frameworks while expanding their functionality to accommodate emerging applications. The modeling framework's dual construction approach—either by assembling circuits or by defining materials and geometries—makes it accessible to diverse research communities, from electrical engineers to material scientists. This flexibility opens new avenues for innovation in chemical sensing, biointerfacing, biomimetic systems, and adaptive materials that exhibit chemical, mechanical, or neuromorphic responses. We anticipate that this tool will serve as a critical enabler for future advancements in these fields, accelerating the design and deployment of soft ionics-based technologies.

\section*{Methods}

 Our model is implemented using the open source and cross domain modeling language, Modelica. Modelica is a domain specific language for programming dynamical systems. There are a number of both academic and commercial implementations of the language, which enjoys support from a wide range of fields. All of the simulation and development of this tool was done in the free and open source tool OpenModelica. The source and results files for our modeling library and each simulation are publicly available at Zenodo \cite{tepermeisterDataCodeHarnessing2024} and on Github. 

\subsection{Indivisible Elements}

Each of the indivisible elements have governing differential algebraic equations that describe their behavior in terms of their material properties and variables defined at their connection with neighboring elements.

\subsubsection{Ground}
The \textit{ground} element provides a voltage reference to the circuits. It has one connector and fixes the voltage to be zero at that connector (Equations \ref{eqn:elements.ground.voltage} and \ref{eqn:elements.ground.flux})

\begin{align}
	\phi &= 0 \label{eqn:elements.ground.voltage}\\
	j_i &= 0 \text{ for $i < n$}\label{eqn:elements.ground.flux}
\end{align}

\noindent where $\phi$ is the electrical potential at the connector and $j_i$ is the flux of species $i$ through the connector. $n$ is the number of mobile species. 

\subsubsection{Infinite bath}
The \textit{infinite bath} element represents an infinite solution with constant concentration. It has one connector and fixes the concentration at that connector (Equation \ref{eqn:elements.bath.concentration}), along with the requirement that no current may flow (Equation \ref{eqn:elements.bath.no_current})
\begin{align}
	C_i &= C^{fixed}_i \text{ for $i < n$} \label{eqn:elements.bath.concentration}\\
	\sum_i j_i z_i &= 0 \label{eqn:elements.bath.no_current}
\end{align}
where $C_i$ is the concentration of species $i$ at the connector, $z_i$ is the valence of species $i$, and $j_i$ is the flux of species $i$ through the connector.
\subsubsection{Volumes}

The \textit{volume} element represents a uniform volume of material and has one connector. It is the main element where mobile species are allowed to accumulate. The governing equation is given in equation \ref{eqn:elements.volume_governing}, which enforces conservation of mass.

\begin{equation}\label{eqn:elements.volume_governing}
	V\frac{dC_i}{dt} = j_i
\end{equation}

\noindent where $V$ is the internal volume of the element, and $j_i$ is the flux of species $i$ through the element's single connector.

\subsubsection{Resistors}

The \textit{resistor} element captures the flow of species in response electrochemical driving forces. It is a two connector element which does not allow ion accumulation, and requires both connectors be made of the same material. We name the two connectors $a$ and $b$. It is governed by equations \ref{eqn:elements.resistors.flux}--\ref{eqn:elements.resistors.upwinding}.

\begin{align}
	j^a_i &= G_i \left(\Delta C_i + \frac{Fz_i \bar{C_i}}{RT}\Delta \phi + \bar{C_i}\Delta\log\gamma_i\right) \label{eqn:elements.resistors.flux}\\
	j^b_i &= -j^a_i\label{eqn:elements.resistors.current_continuity}\\
	\bar C_i &= \begin{cases}
		C^a_i & z_i\Delta \phi > 0\\
		C^b_i & z_i\Delta \phi < 0
	\end{cases} \label{eqn:elements.resistors.upwinding}
\end{align}

\noindent where $G_i$ is the conductance of the element to species $i$, $R$ is the universal gas constant, $T$ is the temperature of the element (assumed constant), $\bar{C_i}$ is the effective concentration of species $i$, $z_i$ is the charge number of that species, $\gamma_i$ is the activity coefficient, and $\phi$ is the electric potential. $\Delta$ represents the difference between the quantities at connector $a$ and connector $b$. $j^{a}_i$ and $j^{b}_i$ represent the fluxes of species $i$ through connectors $a$ and $b$ respectively. The first term in equation \ref{eqn:elements.resistors.flux} represents the concentration gradient driving force, the second term represents the advection of charged mobile species by the electric field, and the third term represents the driving force from differences in activity coefficients. Equation \ref{eqn:elements.resistors.current_continuity} enforces flux continuity in the element, guaranteeing that all the ions that enter one side leave the other. We upwind the effective advected concentration according to equation \ref{eqn:elements.resistors.upwinding}.

\subsubsection{Sharp interface}

The \textit{sharp material interface} element represents a sharp interface between two materials that are locally in electrochemical equilibrium. We neglect the charge accumulation at the interface, using the sharpness assumption to infer that the interface layers themselves store a negligible quantity of species. This equilibrium is described by Equation \ref{eqn:elements.sharp_interface.equilibrium} and---like all elements with no accumulation---Equation \ref{eqn:elements.resistors.current_continuity}.

\begin{equation}\label{eqn:elements.sharp_interface.equilibrium}
	\exp\left({\frac{Fz_i\Delta \phi}{RT}}\right) = \frac{\gamma^a_i C^a_i}{\gamma^b_i C^b_i}
\end{equation}

\subsubsection{Electrodes}
These elements represent the interface between an electrically conductive material and an ionically conductive material. In general, we divide this interface into two regions, an inner Helmholtz plane and a diffuse layer. A detailed discussion of these regions and assumptions are given in Bazant et al. and van Soestbergen et al. \cite{bazantDiffusechargeDynamicsElectrochemical2004, vansoestbergenDiffusechargeEffectsTransient2010}. The electrical potential drop across these two regions combines to produce the potential drop of the overall element as shown in equation \ref{eqn:elements.electrode.potentials_combined}, where $\phi^{d}$ is the voltage drop across the diffuse layer and $\phi^{h}$ is the voltage drop from the electrode surface to the inner Helmholtz plane. Each electrode element differs in how it treats these two layers. In order to connect the flow of charge inside the material with the flow of current inside the electrode conductor, we assume that there is a continuity of current. This continuity is given in equation \ref{eqn:elements.electrode.current_continuity} where $j_i^a$ is the flux of species $i$ in the ionic material and $I$ is the electrical current in the electrically conductive material.

\begin{align}
	\Delta \phi &= \Delta \phi^{d} + \Delta \phi^{h}\label{eqn:elements.electrode.potentials_combined}\\
	I &= -\sum_i F z_i j_i^a \label{eqn:elements.electrode.current_continuity}
\end{align}

\paragraph{Infinite Bath Electrode}

The simplest case is the ideal non-blocking electrode with fixed concentration. In this case, both voltage drops are zero and the concentration is fixed. Under these assumptions, the electrode is governed by equations \ref{eqn:elements.electrode.non-blocking-1}--\ref{eqn:elements.electrode.non-blocking-3}.

\begin{align}
	\phi^d &= 0 \label{eqn:elements.electrode.non-blocking-1}\\ 
	\phi^h &=0 \\
	C_i^a &= C_i^{fixed} \text{ for $i < n$} \label{eqn:elements.electrode.non-blocking-3}
\end{align}

\noindent where $C_i^{fixed}$ is the fixed concentration of species $i$ at the electrode. This is equivalent to the electrode being connected to an infinite bath of species with concentration $C_i^{fixed}$.

\paragraph{Capacitive Blocking Electrode}

For the ideal blocking electrode, no charge is allowed to flow from the electrode surface to the material. Instead, ions accumulating on the inner Helmholtz plane form a capacitor with the electrode surface. We assume that these accumulating ions are in equilibrium with the bulk material, and neglect surface or ion specific contributions to this equilibrium. These assumptions lead to equations  \ref{eqn:elements.electrode.blocking-1}--\ref{eqn:elements.electrode.blocking-3}.

\begin{align}
	\frac{C^*_i}{C^a_i} &= \exp\left({\frac{Fz_i\Delta \phi^d}{RT}}\right) \label{eqn:elements.electrode.blocking-1}\\
	\phi^h &= \frac{Al}{H} \left(F\sum_i C^*_i z_i + \rho_q^a\right)\\
	j_i^a &= Al \frac{d C^*_i}{dt} \label{eqn:elements.electrode.blocking-3}
\end{align}

\noindent where $C^*_i$ is the concentration of species $i$ at the Helmholtz plane, $A$ is the area of the electrode, $l$ is the characteristic length for the inner Helmholtz plane, $H$ is the areal capacitance of the electrode, and $\rho_q^a$ is the fixed charge density of material $a$. Note that $C^*_i$ does not obey the electroneutrality condition. This is a simple model of double layer charging, and while useful for our circuits, many more effects can be included with a more complete double layer model \cite{bazantDiffusechargeDynamicsElectrochemical2004}.

\paragraph{Tafel Electrode}

Finally for the Tafel electrode, we assume that there are $m$ chemical reactions, with the $l$th reaction involving $n_l$ electrons transferring between a species in the at the inner Helmholtz plane and the electrode. We ignore the capacitance of the double layer, and assume that the chemical reaction is exponential in the overpotential. This is described by Equations \ref{eqn:elements.electrode.butler.no_diffuse}--\ref{eqn:elements.electrode.butler.reaction_based_flux}.

\begin{align}
	\phi^d &= 0 \label{eqn:elements.electrode.butler.no_diffuse}\\
	\eta & = \Delta\phi^h - \Delta\phi^\circ \label{eqn:elements.electrode.butler.overpotential}\\
	j_i^a & = -\sum_l \nu_i^l K_l \frac{\prod_k (C_k^a)^{x_k^l}}{\prod_k (C_k^\circ)^{x_k^l}} \exp \left(\frac{\alpha_l n_l F \eta}{RT} \right)\label{eqn:elements.electrode.butler.reaction_based_flux}
\end{align}

\noindent where $\eta$ is the overpotential, $\phi^\circ$ is the equilibrium potential, $\nu_i^l$ is the stoichiometric coefficient of species $i$ in electrochemical reaction $j$, $K_l$ is the rate coefficient for reaction $l$, $C_k^\circ$ is the reference concentration for species $k$, $x_k^l$ is the rate exponent for species $k$ in reaction $l$, $\alpha_l$ is the charge transfer coefficient, and $n_l$ is the number of electrons transferred in the reaction. In addition, we provide an optional current limit $I_{max}$ that caps the total current that flows at very large overpotentials. This extends the applicability of the model without adding significant complexity. 

\subsubsection{Capacitive Material Interface} This element represents a thin dielectric separating two ionic conductors. It has similar equations and assumptions as the \textit{capacitive blocking electrode}, except with the capacitance between the two Helmholtz planes instead of only one. The materials and mobile species do not have to be the same on the two connectors since no ions are allowed to move between them. This is described in equations \ref{eqn:elements.capacitor.voltage_combined}--\ref{eqn:elements.capacitor.fluxes}

\begin{align}
	\Delta \phi &= \Delta \phi^{d_a} + \Delta \phi^{h} - \Delta \phi^{d_b} & \phi^h &= \frac{Q}{H} \label{eqn:elements.capacitor.voltage_combined}\\
	\frac{C^{a*}_i}{C^a_i} &= \exp\left({\frac{2Fz_i\Delta \phi^{d_a}}{RT}}\right) &
	\frac{C^{b*}_k}{C^b_k} &= \exp\left({\frac{2Fz_k\Delta \phi^{d_b}}{RT}}\right)\\
	\frac{Q}{Al} &= \sum_i FC^{a*}_i z_i + \rho_q^{a} & \frac{Q}{Al}&= -\sum_k FC^{b*}_k z_k+ \rho_q^{b}\\
	j_i^a &= Al \frac{d C^{a*}_i}{dt} & j_k^b &= Al \frac{d C^{b*}_k}{dt} \label{eqn:elements.capacitor.fluxes}
\end{align}

\noindent where $C^{a*}_i$, $C^{b*}_k$ is the areal concentration of species $i$ and species $k$ at the Helmholtz plane on side $a$ and side $b$, $A$ is the area of the electrode, $H$ is the areal capacitance of the electrode, $j^{a}_i$ and $j^{b}_k$ represent the fluxes of species $i$ and $k$ through connectors $a$ and $b$ respectively. $\phi^{d_a}$ and $\phi^{d_b}$ represent the diffuse voltage drops on each side, and $\phi^{d_b}$ represents the shared voltage drop between the two Helmholtz planes. $Q$ is the total charge on the capacitor plates. $\rho_q^{a}$ and $\rho_q^{b}$ are the fixed charge densities in the materials on side $a$ and side $b$, and $l$ is a characteristic length for the Helmholtz layer.

\end{multicols}

\printbibliography

\end{document}


\maketitle

\tableofcontents
\listoffigures
\listoftables

\section{Construction of Single Diodes} \label{sec:si.single_diodes}

The trapezoidal diode is formed from thin aqueous polyelectrolytes molded between two thin plates of glass. The polycation is poly(diallyldimethylammonium chloride) (pDADMAC) and the polyanion is poly(2-acrylamido-2-methyl-1-propanesulfonic acid) (pAMPSA), both formed from high molarity solutions of monomers (\qty{5000}{\mol\per\meter\cubed}) \cite{sabbaghDesigningIontronicLogic2023}. Each of the polyelectrolytes is a trapezoid shape with the small sides in contact with the other polyelectrolyte and the larger sides in contact with a \qty{10}{\mol\per\meter\cubed} water bath. In order to represent this geometry using our diode model, we break the trapezoid into two rectangular prisms, one wide and one narrow, with the same total volume as the trapezoid. We choose keep the simple uniform elements instead of creating a trapezoidal element to demonstrate that accurately modeling different diode geometries works well even with simple elements. The rough length and volume of each material is much more important than the specific geometry. We fit the diffusivity ($D_i$) and the fixed charge density($\rho_q$) of each material; We also fit the length of each trapezoid, and the length of the baths. A full table of model parameters is given in Table \ref{tbl:si.diode_parameters}.

The thin film diode is constructed by patterning polyelectrolyte films using photolithography on a Polyethylene terephthalate (PET) sheet. The polyanion is poly(3,4-ethylenedioxythiophene) polystyrene sulfonate (pEDOT:pSS) and the polycation is poly(vinylbenzyl chloride) quaternized by triphenylphospine (PVBPPh3)\cite{gabrielssonPolyphosphoniumBasedBipolar2013}. The geometry of the device matches our model, with both a narrow and wide region of polyelectrolyte films. The main simplification that we make in modeling this device is to replace the overlapping central interface with a butt joint where the ends of the materials touch. For this diode, the thicknesses of the two polyelectrolyte regions are fit in addition to $D_i$ and $\rho_q$ for each material. A full table of model parameters is given in Table \ref{tbl:si.diode_parameters}. Some additional details of this geometry not described in the original paper were obtained in personal communications with the authors. 

\section{Resistor Prism}

This element behaves like a resistor, but calculates $G_i$, the conductance, from the given geometry as

\begin{equation}
	G_i = \frac{A \bar{D}_i}{L}
\end{equation} 

\noindent where $A$ is the area of the element, $L$ is the length, and $\bar{D_i}$ is the average of the diffusivity for species $i$ in that material calculated from the material state at connector $a$ and connector $b$.

\section{Uniform Material Prism} \label{sec:si.uniform_material_prism}

This element that combines \textit{volume}s and \textit{resistor}s to represent a block of a single material. This prism is broken into basic elements based on 1d finite volumes. Each \textit{volume} element represents a region of the material, while the \textit{resistor}s have a conductance calculated from the resistance of the materials connecting the center of the volumes. The number of \textit{volume}s is a tunable parameter. In order to approximate the continuum time series response as closely as possible, the elements closer to the edges of the prism are thinner .

\begin{figure}
	\centering
	\includegraphics[width=\textwidth]{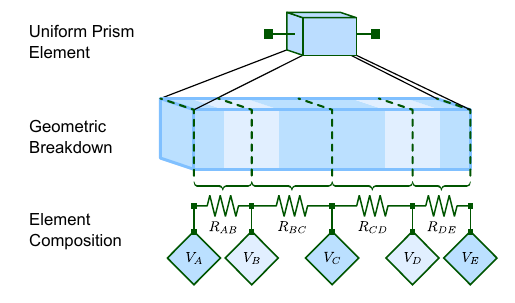}
	\caption[Uniform material prism construction]{The structure of the \textit{uniform material prism} element, showing how it is broken down into indivisible elements. The symbol for the \textit{uniform material prism} is shown at the top. In the middle is a geometric breakdown of the element, showing how it is partitioned into different regions. The light and dark blue regions indicate the location of each volumes, while the green dotted lines indicate the endpoints of each of the resistors. The color of each volume matches the volume elements that are shown in the last row. The last row shows the specific indivisible elements that compose the prism.}
	\label{fig:si.geometric_to_elements}
\end{figure}

A diagram of the \textit{uniform material prism} is shown in Figure \ref{fig:si.geometric_to_elements}. The length of each of the volumes is determined as given in Equations \ref{eqn:si.geometric.unnormalized_length_factor} and \ref{eqn:si.geometric.lengths}.

\begin{align}
	\lambda_i &= \frac{1}{e^{\frac{-ki}{n}}+ e^{-k\left(1-\frac{i}{n}\right)}} \label{eqn:si.geometric.unnormalized_length_factor}\\
	l_i &= \frac{L\lambda_i}{\sum_{i=1}^n \lambda_i}\label{eqn:si.geometric.lengths}
\end{align}

\noindent where $k$ is a sharpness factor, $n$ is the number of sub-volumes, $L$ is the total length of the element, and $l_i$ is the length of the $i$th sub-volume. All the sub-volumes share an area with the high level element. The resistors have lengths that span the distances between the centers of adjacent volumes, except at the edge where the resistors have a length that goes all the way to the edge.

\section{Tables and Figures}


\begin{table}\centering
	\small
	\sisetup{round-mode=figures,round-precision=2, exponent-mode=engineering}
	\rowcolors{2}{gray!10}{white}
	\caption[Single diode properties]{Properties for the elements used to make the single diodes. The red numbers are fit to the experimental data using a least squares fitting procedure with a soft l1 loss function. The numbers in blue are calculated based on the fit numbers and details known from the literature. The numbers in black are directly estimated or taken from the experimental literature. In the material properties sub-table, $D$ Multiplier is a constant that is multiplied by the pure water diffusivities \cite{TableDiffusionCoefficients2020} of each mobile species to calculate its diffusivity in the material.}
	\subcaption*{Element Properties}
	\begin{tabular}{l l l S S S[round-mode=places,round-precision=0] S[round-mode=places,round-precision=0]}
		{Diode} & {Element}  & {Material} & {Length (\unit{\milli\meter})} & {Area (\unit{\milli\meter\squared})} & \shortstack{Number of\\ segments, $n$} & {Sharpness, $k$}\\
		\midrule
		
		\cellcolor{white}& bath A            & water & \color{DarkRed} 4.200 & 1.5e-1 & 5 & 10\\
		\cellcolor{white}& wide polyanion    & pAMPSA & \color{DarkRed}.110  & \color{DarkBlue}2.575e-2 & 5 & 10\\
		\cellcolor{white}& narrow polyanion  & pAMPSA & \color{DarkBlue}.110  & \color{DarkBlue}2.2e-2 & 5 & 10\\
		\cellcolor{white}& wide polycation   & pDADMA & \color{DarkRed}.500  & \color{DarkBlue}2.875e-2 & 5 & 10\\
		\cellcolor{white}& narrow polycation & pDADMA & \color{DarkBlue}.500  & \color{DarkBlue}4.65e-2 & 5 & 10\\
		\cellcolor{white}\multirow{-6}{*}{Sabbagh et al.}      & bath b            & water & \color{DarkBlue} 4.200 & 1.5e-1 & 5 & 10\\
		
		\cellcolor{gray!10}& bath A            & water & 1.2 & 1.2e-1 & 5 & 10\\
		\cellcolor{gray!10}& wide polyanion    & pEDOT pSS & 0.375  & \color{DarkRed}1.8e-5 & 5 & 10\\
		\cellcolor{gray!10}& narrow polyanion  & pEDOT pSS & .10  & \color{DarkBlue}0.9e-5 & 5 & 10\\
		\cellcolor{gray!10}& wide polycation   & pVBPPh3 & .100  & \color{DarkRed}8.46e-4 & 5 & 10\\
		\cellcolor{gray!10}& narrow polycation & pVBPPh3 & .375  & \color{DarkBlue}4.23e-4 & 5 & 10\\
		\cellcolor{gray!10}\multirow{-6}{*}{Gabrielsson et al.}      & bath b            & water & 1.2 & 1.2e-1 & 5 & 10
		
	\end{tabular}
	\bigskip
	\subcaption*{Material Properties}
	\begin{tabular}{l S[exponent-mode=threshold] S[exponent-mode=threshold,exponent-thresholds = -4:4]}
		{Material} & {$D$ Multiplier}  & {$\rho_q/F$}\\
		\midrule
		water & 1.0 & 0\\
		pAMPSA & \color{DarkRed}0.14 & \color{DarkRed}-0.68e3\\
		pDADMA & \color{DarkRed}0.2 & \color{DarkRed}1.05e3\\
		pEDOT pSS & \color{DarkRed}0.426 & -1e3\\
		pVBPPh3 & \color{DarkRed}3.19e-3 & 3e3\\
		
	\end{tabular}
	\label{tbl:si.diode_parameters}
\end{table}

\begin{figure}
	\centering
	\includegraphics[width=\textwidth]{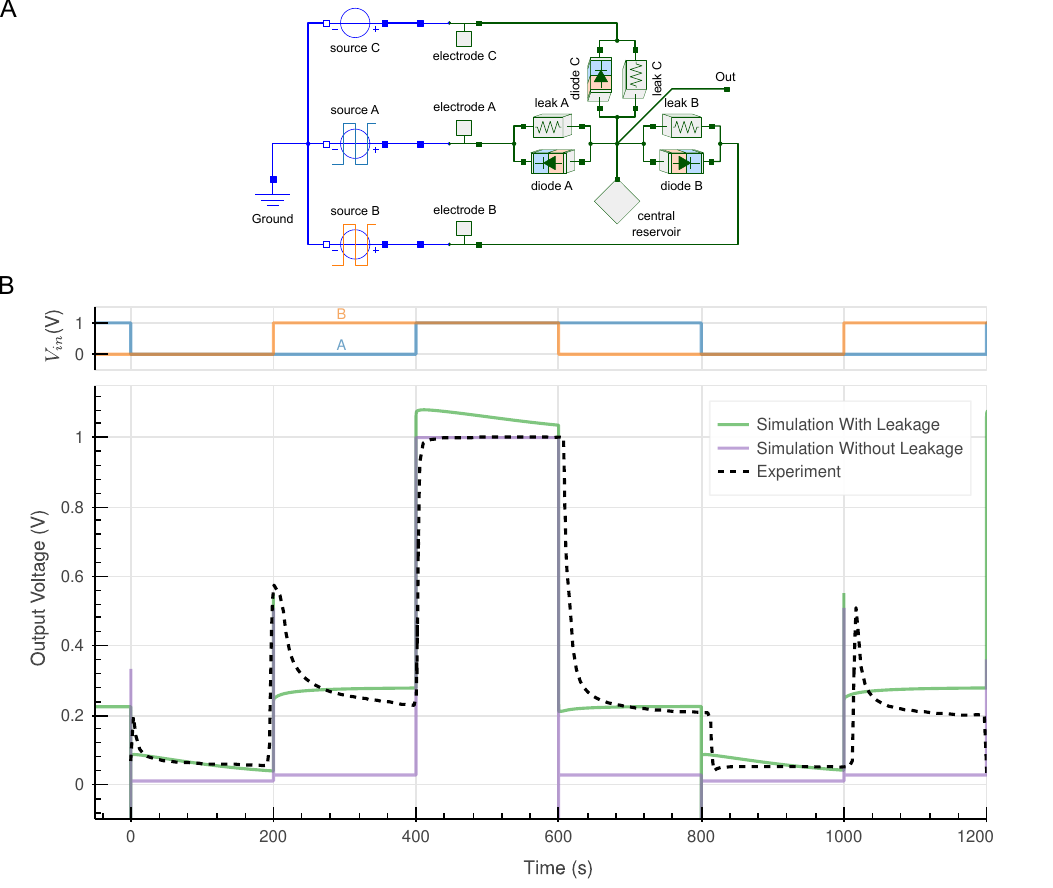}
	\caption[\textsc{and} gate leakage]{Comparison of the \textsc{and} gate with and without leaking diodes. A) Diagram of the \textsc{and} gate with added leakage. B) Input voltage vs time for the power supplies connected to diode A and diode B (upper). The voltage sweeps through all possible binary inputs. Comparison of the gate output voltage vs time (upper) among the simulation (dashed), gate with leakage (green), and gate without leakage (purple). The resistances for each of the leakage resistors are fit to match the experimental data.}
	\label{fig:si.logic_gate_leakage}
\end{figure}

\begin{figure}
	\centering
	\includegraphics[width=\textwidth]{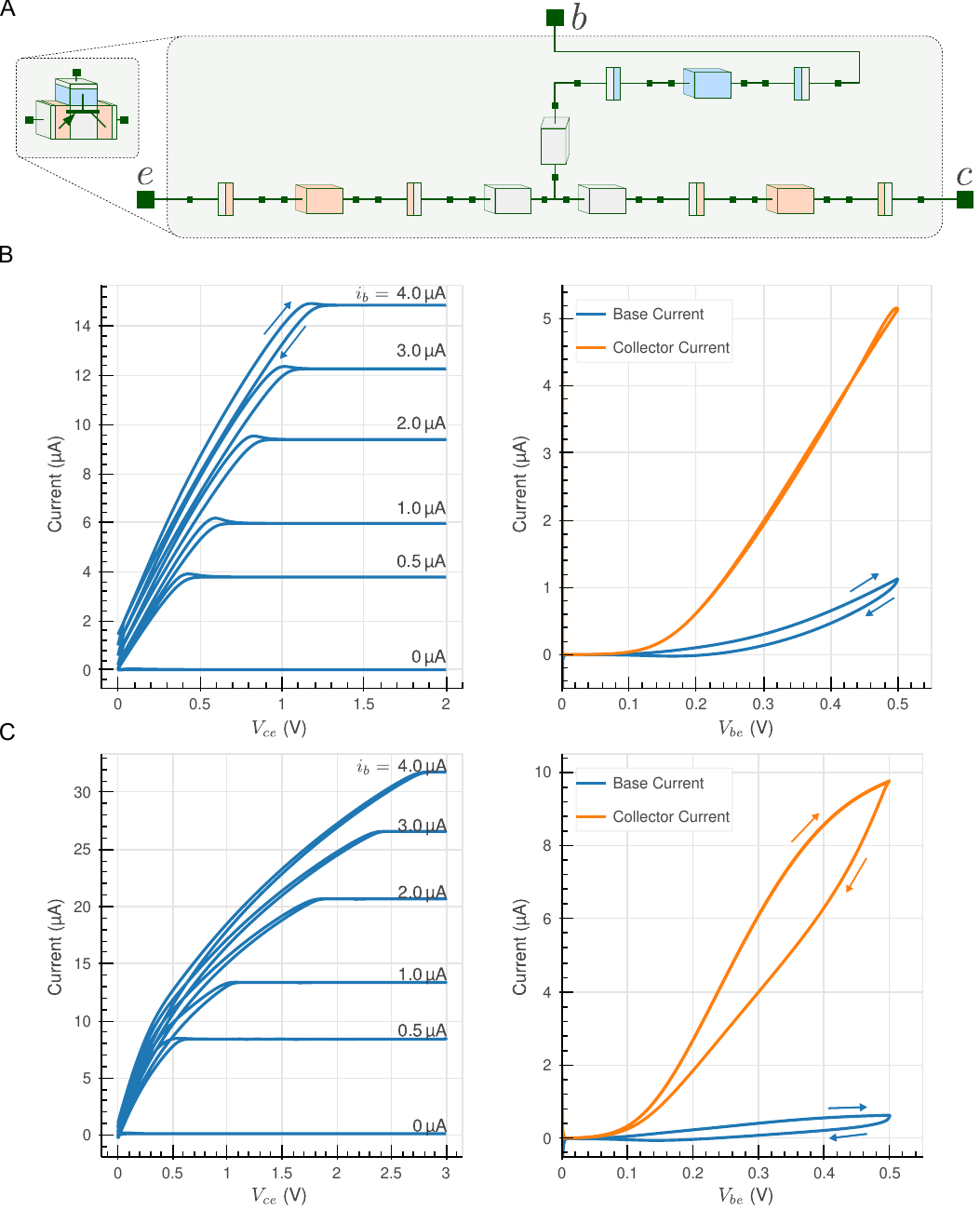}
	\caption[Transistor element response]{Properties of the transistor elements used in the robot control circuit. A) The elements and their connections. B) The transistor output characteristics (left) and currents vs base voltage (right) for the oscillator transistors. The voltage was scanned at \qty{4}{\milli\volt\per\second} C) The transistor output characteristics (left) and currents vs base voltage (right) for the actuator transistors. The voltage was scanned at a slower \qty{1}{\milli\volt\per\second} since the response was slower.}
	\label{fig:si.transistor}
\end{figure}

\begin{figure}
	\centering
	\includegraphics[width=\textwidth]{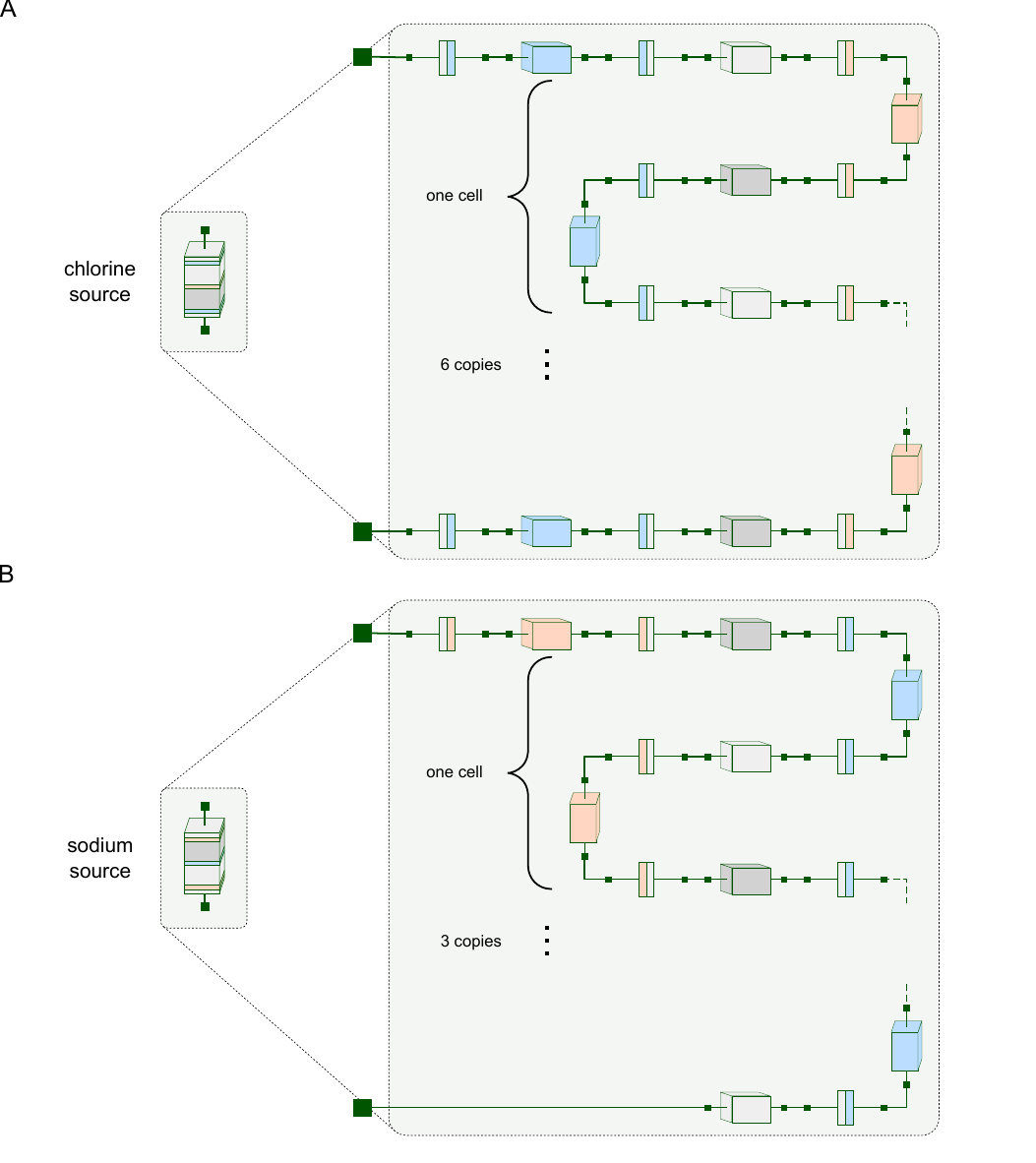}
	\caption[Electrodialysis structure]{Structure of the electrodialysis devices used in the robot control circuit, showing the elements that it is made from. The darker grey materials have high salt concentration initially, while the light grey materials have low salt concentration. The energy associated with this concentration imbalance powers the robot control circuit. A) The chlorine source. B) The sodium source}
	\label{fig:si.electrodialysis}
\end{figure}

\begin{figure}
	\centering
	\includegraphics[width=\textwidth]{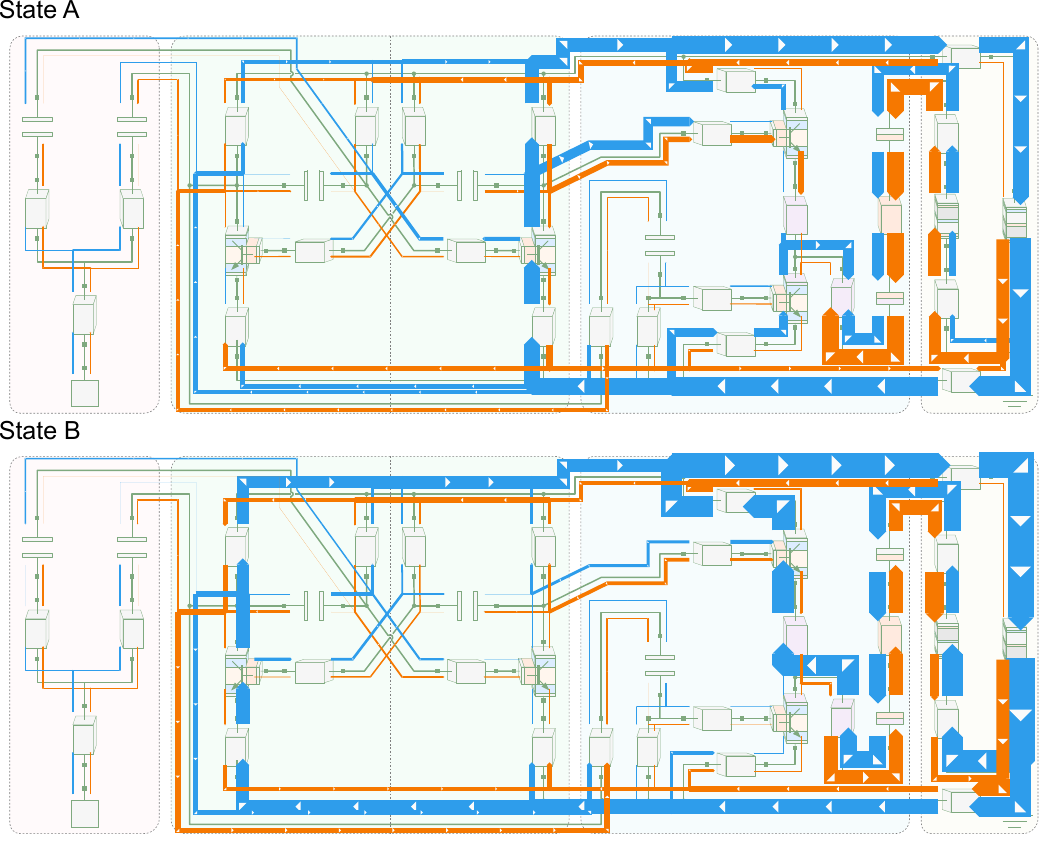}
	\caption[Robot circuit ion flows]{Flows of sodium (orange) and chlorine (blue) in the robot circuit in the two oscillator states. State A is when the right oscillator transistor is on and the actuator is being enriched with salt. State B is when the left oscillator transistor is on and the actuator is being depleted of salts. The width of the lines is proportional to the square root of the flux of each species in order for smaller fluxes to be visible. The white arrows and line arrowheads point in the direction of the flux for each species.}
	\label{fig:si.oscillator_circuit.currents}
\end{figure}

\begin{figure}
	\centering
	\includegraphics[width=\textwidth]{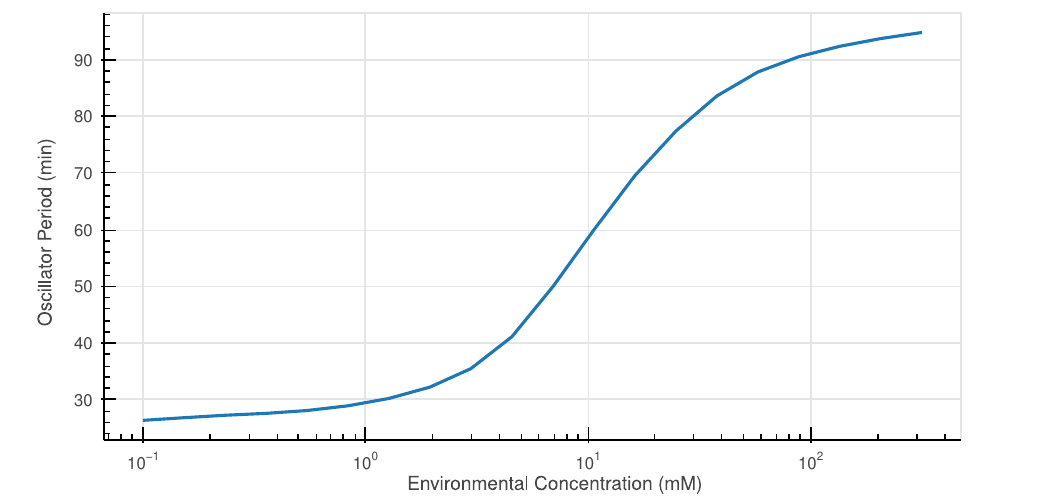}
	\caption[Robot circuit sensor-oscillator response]{The response curve for the oscillation period of the robot control circuit oscillator vs the concentration of salt in the external environment measured by the sensor module.}
	\label{fig:si.oscillator_circuit.sensor_sensitivity}
\end{figure}

\begin{figure}
	\centering
	\includegraphics[width=\textwidth]{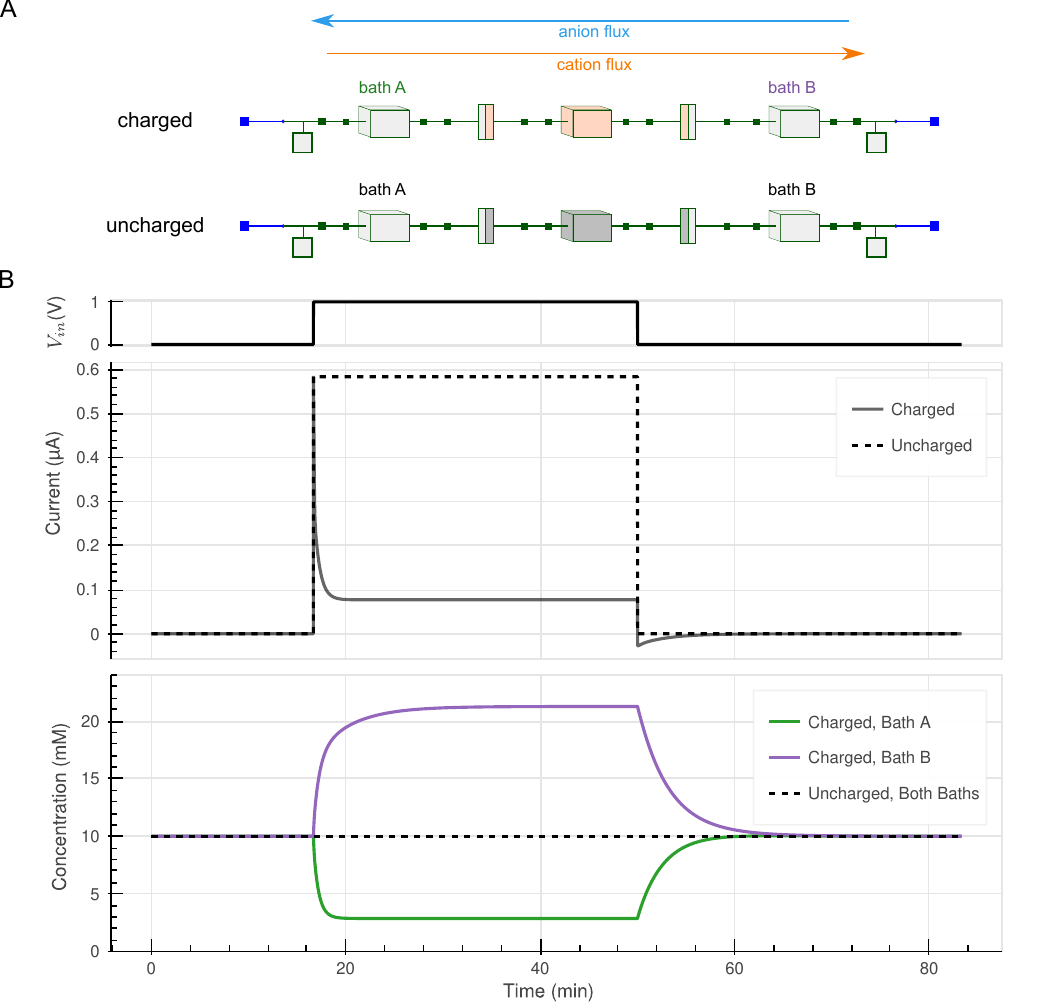}
	\caption[Wire enrichment effects]{Demonstration of the enrichment and depletion effect for neutral wires near polyelectrolytes. A) Two diagrams depicting the elements used for the demonstration. The upper model has a charged polyanion in the center, while the lower model has a neutral polymer. B) Input voltage applied to the electrodes vs time (upper), with positive on the left and negative on the right. Current vs time (middle) showing the case with the charged central polymer (solid) and the uncharged central polymer (dotted). The resistance of the uncharged central polymer has been chosen to match the initial current of the charged case. The average salt concentration in the baths vs time (lower). With the charged polymer in the center, the concentration in bath A (green) increases and the concentration in bath B (purple) decreases while current is flowing through the wire. When the voltage is switched off, the concentrations return to nominal. When the polymer in the center is not charged (dotted), the concentration of salt in both baths remains constant.}
	\label{fig:si.wire_concentration_changes}
\end{figure}

\begin{figure}
	\centering
	\includegraphics[width=\textwidth]{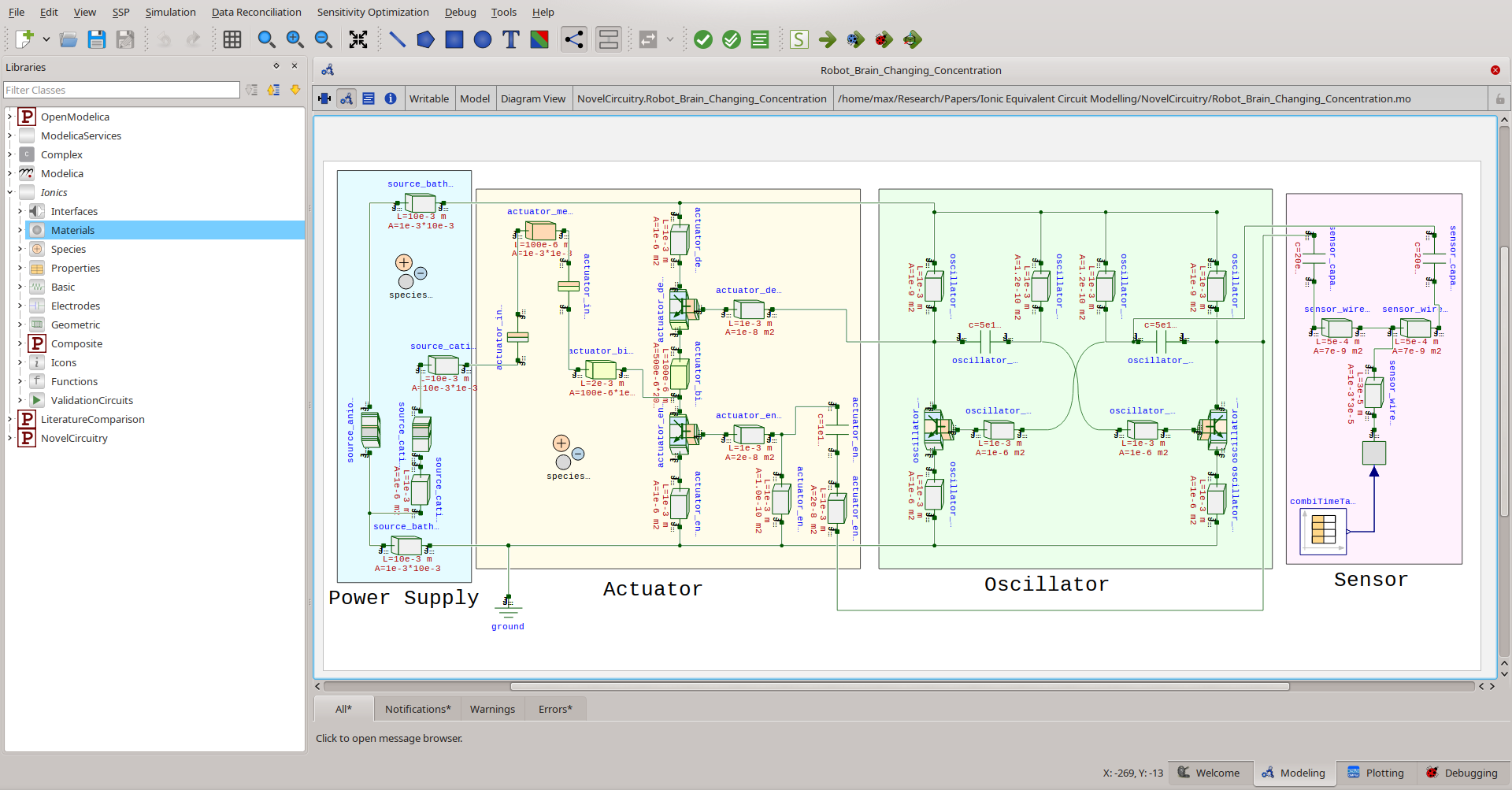}
	\caption[Tool screenshot]{Screenshot of OMEdit, the open source graphical lumped element design tool that we used to develop all of the elements and circuitry in this paper \cite{OMEditOpenModelicaConnection2023}. In order to interact with these elements through a GUI, we have drawn graphical representations of each element that dynamically show what material they are made from and some key properties of the element. This makes designing circuits easy without writing code.}
	\label{fig:si.tool_screenshot}
\end{figure}

\begin{table}\centering
	\fontsize{7pt}{11pt}
	\selectfont
	\sisetup{round-mode=places,round-precision=3}
	\caption[Computational expenses]{Computational expenses for each simulation run in this paper. Each simulation was run on a single Xeon E3-1225 v6 CPU. The time ratio in the last column indicates how much faster than real-time the simulation process was, excluding translation and compilation time. The number of steps is the minimum number of timesteps taken by the integrator, but for some simulations many sub-steps are taken within each time-step. All the simulations use the DASSL integrator \cite{petzoldDescriptionDASSLDifferential1982} and symbolic jacobians.}
	\rowcolors{2}{gray!10}{white}
	
	\begin{tabular}{l l S[round-precision=0] S[round-precision=0] S S S S S[round-precision=0]}
		{Simulation Name} & {Variation}  & \shortstack{Simulated \\Time (s)} & \shortstack{Simulated \\Step Count} & \shortstack{Translation \\Wall Time (s)} & \shortstack{Compilation \\Wall Time (s)} & \shortstack{Simulation \\Wall Time (s)} & \shortstack{Total \\Wall Time (s)} & \shortstack{Time Ratio} \\
		\midrule
		\cellcolor{white}& \qty{5}{\second} & 275.000000 & 100 & 1.909586 & 2.836057 & 1.150377 & 5.897824 & 239.052104 \\
		\cellcolor{white}& \qty{10}{\second} & 280.000000 & 100 & 1.740870 & 2.285863 & 1.072501 & 5.099373 & 261.071969 \\ 
		\cellcolor{white} & \qty{20}{\second} & 290.000000 & 100 & 2.097819 & 2.927844 & 1.173430 & 6.199245 & 247.138760 \\
		\cellcolor{white} & \qty{30}{\second} & 300.000000 & 100 & 1.743431 & 2.722086 & 1.118045 & 5.583710 & 268.325501 \\
		\cellcolor{white}& \qty{60}{\second} & 330.000000 & 100 & 2.082351 & 2.471515 & 1.101211 & 5.655233 & 299.670103 \\
		\cellcolor{white}\multirow{-6}{*}{Berggren single diode} & \qty{90}{\second} & 360.000000 & 100 & 1.942315 & 2.443866 & 1.277505 & 5.663860 & 281.799243 \\
		
		Berggren full rectifier &  & 1500.000000 & 4000 & 10.998105 & 10.354323 & 0.241581 & 21.594220 & 6209.105409 \\
		
		Yossifon single diode&  & 6000 & 3998 & 2.042535 & 3.834310 & 2.174754 & 8.051748 & 2758.932866 \\
		
		& no leakage & 1200.000000 & 4000 & 6.260556 & 6.724027 & 5.608815 & 18.593550 & 213.948943 \\
		
		\cellcolor{white}\multirow{-2}{*}{Yossifon \textsc{and} gate} & leakage & 2400.000000 & 4000 & 6.883869 & 7.598949 & 17.570704 & 32.053710 & 136.590998 \\
		\cellcolor{gray!10} robot controller &  & 36000 & 32000 & 30.818574 & 20.394211 & 1679.471439 & 1730.684509 & 21.435315 \\
		
		\cellcolor{white}  & trace  & 6000.000000 & 60000 & 2.123263 & 2.920194 & 4.796688 & 9.840386 & 1250.862984 \\
		
		\cellcolor{white}\multirow{-2}{*}{oscillator transistor}  &  amplification& 12000.000000 & 12000 & 2.370988 & 2.602857 & 3.118822 & 8.093119 & 3847.606387 \\
		
		\cellcolor{gray!10}& trace & 36000.000000 & 36000 & 2.041085 & 3.385338 & 16.288105 & 21.716443 & 2210.201805 \\
		
		\cellcolor{gray!10}\multirow{-2}{*}{actuator transistor} & amplification & 36000.000000 & 36000 & 2.041085 & 3.385338 & 16.288105 & 21.716443 & 2210.201805 \\
		
		\cellcolor{white} & charged & 5000.000000 & 5000 & 1.113252 & 1.922761 & 0.204792 & 3.240978 & 24415.017642 \\
		
		\cellcolor{white}\multirow{-2}{*}{wire enrichment} & uncharged & 5000.000000 & 5000 & 1.036089 & 1.464619 & 0.052397 & 2.553235 & 95425.662102 \\
		
	\end{tabular}
	\label{tbl:si.computational_cost}
\end{table}

\clearpage
\printbibliography